\documentclass{aa}  
\usepackage{graphicx}
\usepackage[varg]{txfonts}
\usepackage{natbib}
\bibpunct{(}{)}{;}{a}{}{,}
\begin{document}
\title{Assessing the reliability of Friends-of-Friends groups \\
on the future Javalambre Physics of the \\
Accelerating Universe Astrophysical Survey}
\author{A. Zandivarez\inst{1,2,3}\fnmsep\thanks{arielz77@gmail.com} \and 
E. D\'iaz-Gim\'enez\inst{1,2,3} \and 
C. Mendes de Oliveira\inst{3} \and 
B. Ascaso\inst{4} \and
N. Ben\'itez\inst{4} \and \\ 
R. Dupke\inst{5,6} \and
L. Sodr\'e Jr.\inst{3} \and
J. Irwin\inst{7} 
}

\institute{
Instituto de Astronom\'{\i}a Te\'orica y Experimental, IATE, CONICET, C\'ordoba, Argentina
\and
Observatorio Astron\'omico, Universidad Nacional de C\'ordoba, Laprida 854, X5000BGR, 
C\'ordoba, Argentina
\and 
Instituto de Astronom\'ia, Geof\'isica e Ciencias Atmosfericas, IAG, USP, Rua do Mat\~ao 1226,
S\~ao Paulo, Brazil
\and
Instituto de Astrofisica de Andaluc\'ia (CSIC), Apdo. 3044, 18008 Granada, Spain
\and
University of Michigan, Ann Arbor, MI 48109, USA; Eureka Scientific Inc., Oakland, CA 94602-3017, USA
\and
Observatório Nacional, Rua Gal. José Cristino, 20921-400 Rio de Janeiro, Brazil
\and
Department of Physics and Astronomy, University of Alabama, Box 870324, Tuscaloosa, AL 35487, USA
}
\date{Received XXX; accepted XXX}
\abstract{}
{We have performed a detailed analysis of the ability of the friends-of-friends algorithm
in identifying real galaxy systems in deep surveys such as the future Javalambre Physics of the Accelerating 
Universe Astrophysical Survey. Our approach is two-fold, i.e., assessing the reliability
of the algorithm in both real and redshift space. In the latter, our intention is also to determine
the degree of accuracy that could be achieved when using spectroscopic or photometric 
redshift determinations as a distance indicator.}
{We have built a light-cone mock catalogue using synthetic galaxies constructed
from the Millennium Run Simulation I plus a semi-analytical model of galaxy formation.
We have explored different ways to define the proper linking length parameters of the algorithm
in order to perform an identification of galaxy groups as suitable as possible in each case.}
{We find that, when identifying systems in redshift space using spectroscopic information, the 
linking lengths should take into account the variation of the luminosity function with redshift as 
well as the linear redshift dependence of the 
radial fiducial velocity in the line of sight direction. When testing purity and completeness of the group
samples, we find that the best resulting group sample reaches values of $\sim 40\%$ and $\sim 70\%$ of systems 
with high levels of purity and completeness, respectively, when using spectroscopic information. 
When identifying systems using photometric redshifts, we adopted a probabilistic approach to link galaxies 
in the line of sight direction. Our result suggests that it is possible to identify a sample of groups with 
less than $\sim 40\%$ false identification at the same time as we recover around $60\%$ of the true groups.}
{This modified version of the algorithm can be applied to deep surveys provided that 
the linking lengths are selected appropriately for the science to be done with the data.}

\titlerunning{Assessing the reliability of FoF groups on the future J-PAS}
\maketitle
\section{Introduction} 
The study of galaxy systems is one of the most important topics of extragalactic astronomy
because the history of galaxy formation and evolution is encrypted in these 
density peaks. Analysing the properties of galaxies in groups at different times is a 
direct probe of how the local environment shapes the galaxies inside of them, offering a direct
insight into the physics that has occurred within the halos.

In order to use these great laboratories to improve our understanding of the Universe, it 
is crucial to define them properly. Then, it is necessary to implement an identification 
criterion to define galaxy systems. Through the decades, defining the proper algorithm to 
identify galaxy systems has troubled scientists.
Many attempts have been carried out on the search of the most suitable method to identify 
galaxy systems using optical properties (see \citealt{gal06} for a review of different identification methods). 
Among them, we can highlight the following: 
methods that use positional information of galaxies to detect density peaks over a background density 
\citep[e.g.][]{couch91,dalton97,ramella01,merchan02,trevese07,gillis11,farrens11}; methods that
include some observational restrictions for a given type of galaxy, like their colours, 
magnitudes and their membership to a red sequence \citep[e.g.][]{gladders00,goto02,miller05,koester07};
and finally, methods that model cluster properties such as luminosity and density profiles through different
probability approaches \citep[e.g.][]{shectman85,postman97,kepner99,gal00,milke10,ascaso12}.

Among all these different methods, those based only on the geometric positional information of 
galaxies have the advantage that they do not bin the data or impose any constraints 
on the physical properties of the systems to avoid selection biases.
The most extensively used finding algorithm that follow this criterion is the Friends-of-Friends (FoF)
technique, which detects density enhancements in 3-dimensions by searching galaxy pairs that 
are closer than a given separation. When applied to an observational catalogue, the FoF algorithm 
makes use of the angular coordinates and the spectroscopic redshifts of the galaxies.  
Nevertheless, to identify groups in redshift space one has to deal with certain difficulties.
One of them is the fact that in most cases the observational samples are flux limited, for which 
the observed decreasing galaxy number density as a function of redshift should be taken into account.
Another important issue is the peculiar velocities of galaxies in groups, since they
elongate groups in the redshift (line-of-sight) direction making them harder to detect, 
and may cause group members to be linked with field galaxies or even to merge 
into another group. Although the FoF technique has been widely used to find groups and clusters in 
galaxy surveys, it has not been tested properly at intermediate and 
high redshifts. Therefore, it is very important to put the method under a highly detailed testing
process to clearly determine its capability to recover real systems.   

In the last years, several medium band photometric surveys (e.g., COMBO-17 - \citealt{wolf04}, 
COSMOS 21 - \citealt{ilbert09}, ALHAMBRA Survey - \citealt{moles08,molino13}, SHARDs - \citealt{perez13}) 
have become available. These surveys provide $\sim 1\%$ photometric redshift resolution, 
and very valuable datasets to identify galaxy concentrations. Future surveys 
will provide hundreds of millions of galaxies with this photo-z resolution,
make specially important to study and develop the application of FoF algorithms to photometric redshift
datasets, a task which is not straightforward, due to the pronounced blurring of galaxy systems in 
redshift space and the sometimes complex shape of the photometric redshift error distributions. 
There are several authors that have proposed a modified FoF algorithm to be applied
to photometric surveys \citep[e.g.][]{botzler04,liu08,li08,vanbreu09}.
Beyond the chosen method, all the parameters and scaling relations of any algorithm
should be carefully tested in order to apply one of these methods on a given deep
photometric survey. 

One of the most promising international projects to build a wide field 
photometric survey is the Javalambre Physics of the Accelerating Universe Astrophysical 
Survey (J-PAS\footnote{jpas.org}, \citealt{benitez09}; Benitez et al. 2013, in prep.) 
which will cover more than 8000 square degrees 
in 54 narrow bands and 5 broad bands in the optical frequency range. 
The survey, which is an international collaboration mainly between
Spain and Brazil, will be carried out using two telescopes of 2.5 m and 0.8 m apertures,
which are being built at Sierra de Javalambre, in Spain \citep{benitez09,moles10}. 
The catalogue is planned to take 4-5 years to be undertaken and it is expected to map
$8,000$ $deg^2$ down to an apparent magnitude of $i_{AB} \sim 23$. 

The advent of deep photometric surveys with reliable estimates of 
photometric redshifts, as the future J-PAS, will demand a well-tailored set of 
tools in order to perform different statistical studies. 
Among them, the availability of different algorithms to
extract reliable samples of galaxy systems is quite important. 
However, in order to test the different observational restrictions in the 
identification procedure, we must use reliable mock 
galaxy catalogues built from cosmological numerical simulation 
with all the 3-d positional information. One of the largest cosmological numerical 
simulations is the Millennium Simulation \citep{springel05}. 
When combined with semi-analytic models of galaxy formation, this simulation 
constitutes a very useful tool to mimic the observational constraints of a given 
catalogue under study. The several snapshots available for this numerical
simulation at different times can allow the construction of very detailed light-cone 
mock catalogues that include the corresponding effects of galaxy evolution up to redshift values
similar to those expected to be achieved with the future J-PAS ($z\sim 1$).

The aim of this work is performing a detailed analysis of the capability of a 
modified FOF algorithm to identify galaxy systems in a deep photometric redshift survey like the
future J-PAS. The adopted modified FoF algorithm is the one developed by \cite{liu08}, 
known as Probability FoF. This method uses a probability distribution function to model
the photometric redshift uncertainties, obtaining a very realistic way to deal with the 
radial linking length without introducing artificial slices in the survey.  
Our work involves testing each observational restriction in order
to disentangle possible problems introduced in the identification process. 
This task is performed on a J-PAS light-cone mock galaxy catalogue constructed using the 
semi-analytical galaxies extracted from the Millennium Simulation \citep{guo11}.   
Our study intends to determine the purity and completeness of a resulting galaxy group sample 
obtained from a group identification algorithm that only uses $2\frac{1}{2}$ 
(angular coordinates+redshifts) positional 
galaxy information and the usefulness of this sample to become an input catalogue for further 
refinements adding other observational properties. 

The layout of this paper is as follows: in section 2, we describe the N-body 
simulation and the semi-analytic model of galaxy formation used to 
build the mock catalogue. In section 3 we describe the implementation
of the FoF algorithm and the modifications needed in order to identify
groups in deep redshift surveys as well as photometric ones. We also include 
in this section the percentage of purity and completeness of the resulting
finder algorithm as a function of redshift. Finally, in section 4 we 
summarise our results and discuss the statistical implications of using this 
type of algorithm in deep photometric surveys.

\section{The mock catalogue}
We build a light-cone mock catalogue using a simulated set of galaxies extracted
from the \cite{guo11} semi-analytic model of galaxy formation applied on 
top of the Millennium Run Simulation I. 

\subsection{The N-body simulation}
The Millennium Simulation is a cosmological Tree-Particle-Mesh 
\citep{xu95} N-body Simulation \citep{springel05}, which evolves 
10 billion ($2160^3$) dark matter particles in a 500 $h^{-1} \ Mpc$ periodic 
box, using a comoving softening length of 5 $h^{-1} \ kpc$. The cosmological 
parameters of this simulation are consistent with WMAP1 data \citep{spergel03}, 
i.e., a flat cosmological model with a 
non-vanishing cosmological constant ($\Lambda$$CDM$): $\Omega_m$ =0.25, 
$\Omega_b$=0.045, $\Omega_{\Lambda}$=0.75, $\sigma_8$=0.9, $n$=1 and $h$=0.73. 
The simulation was started at $z$=127, with the particles initially positioned 
in a glass-like distribution according to the $\Lambda$$CDM$ primordial 
density fluctuation power spectrum. The $10^{10}$ particles of mass 
$8.6\times10^8 h^{-1} \ M_{\odot}$ are then advanced with the TPM code, 
using 11,000 internal time-steps, on a 512-processor supercomputer. 
The full particle data (positions and velocities) between $z=20$ and $z=0$ 
were stored at 60 output times spaced in expansion factor according to
$\log(1+z_i)=i(i+35)/4200$. Additional outputs were added at $z$ = 30, 
50, 80, 127 to produce a total of 64 snapshots in all. 

\subsection{The semi-analytic model}
In order to obtain a simulated galaxy set we adopt the \cite{guo11} semi-analytic 
model, which fixed several open issues present in some of its predecessors. 
For instance, the authors increased the efficiency of supernova feedback by 
introducing a direct dependence of the amount of gas reheated and ejected on 
the virial mass of the host halo. Although the resulting model fits 
the stellar mass function of galaxies well at low redshifts, it still over produces 
low-mass galaxies at $z>1$. \cite{guo11} also introduced a more realistic treatment 
of satellite galaxy evolution and of mergers, allowing satellites to continue 
forming stars for a longer period of time and reducing the satellite 
excessively rapid reddening. The model also includes a treatment of the 
tidal disruption of satellite galaxies.

This model produces a complete sample when considering galaxies with rest 
frame absolute magnitude in the SDSS $i$-band brighter than -16.4, which 
implies galaxies with stellar masses larger than $\sim 10^8 h^{-1} \ M_{\odot}$.

Since different cosmological parameters have been found from WMAP7 \citep{komatsu11}, 
one may argue that the studies carried out in the present simulation could produce results 
that do not agree with the current cosmological model. However, \cite{guo13} have 
demonstrated that the abundance and clustering of dark halos and galaxy properties, 
including clustering, in WMAP7 are very similar to those found in WMAP1 for $z\leq3$, 
which is the redshift range of interest in this work (see Sect.~\ref{mock}).

\subsection{Mock catalogue construction}
\label{mock}
\begin{figure}
\begin{center}
\includegraphics[width=80mm]{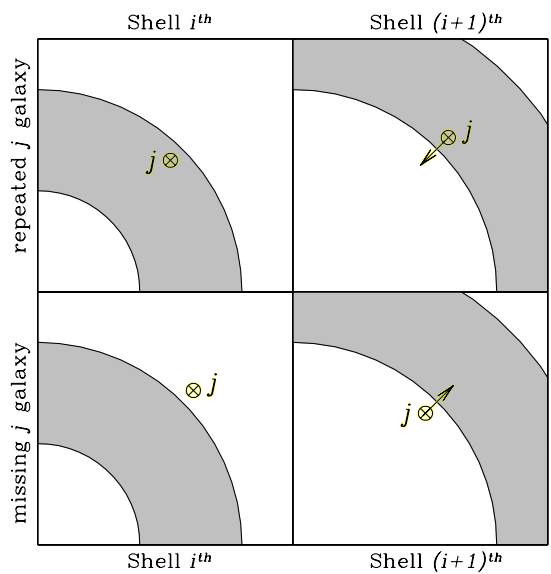}
\caption{
Illustration of the process of galaxies being included twice or none
in the mock catalogue for two consecutive snapshots.
\emph{Upper panels}: The case when a galaxy is 
counted twice when constructing a light-cone using different snapshots. The
\emph{grey} region shows the snapshot under consideration. The \emph{right panel} 
shows the galaxy $j$ in a previous time, inside the $i+1^{th}$ shell, showing the direction
of movement of the galaxy. Due to this direction of movement and the width of the
shell, the galaxy $j$ will appear also inside the $i^{th}$ shell of an earlier time 
(\emph{left panel}), and consequently, being included twice. \emph{Lower panels}: 
The case when a galaxy is missing when constructing a light-cone. The \emph{right panel} shows 
the galaxy $j$ in a previous time, outside the $i+1^{th}$ shell, and showing its direction
of movement. Due to this situation, in an earlier time, galaxy $j$ will also appear
outside the $i^{th}$ shell (\emph{left panel}), resulting in a missing galaxy in both shells.
}
\label{fig1}
\end{center}
\end{figure}
\begin{figure*}
\begin{center}
\includegraphics[width=\hsize]{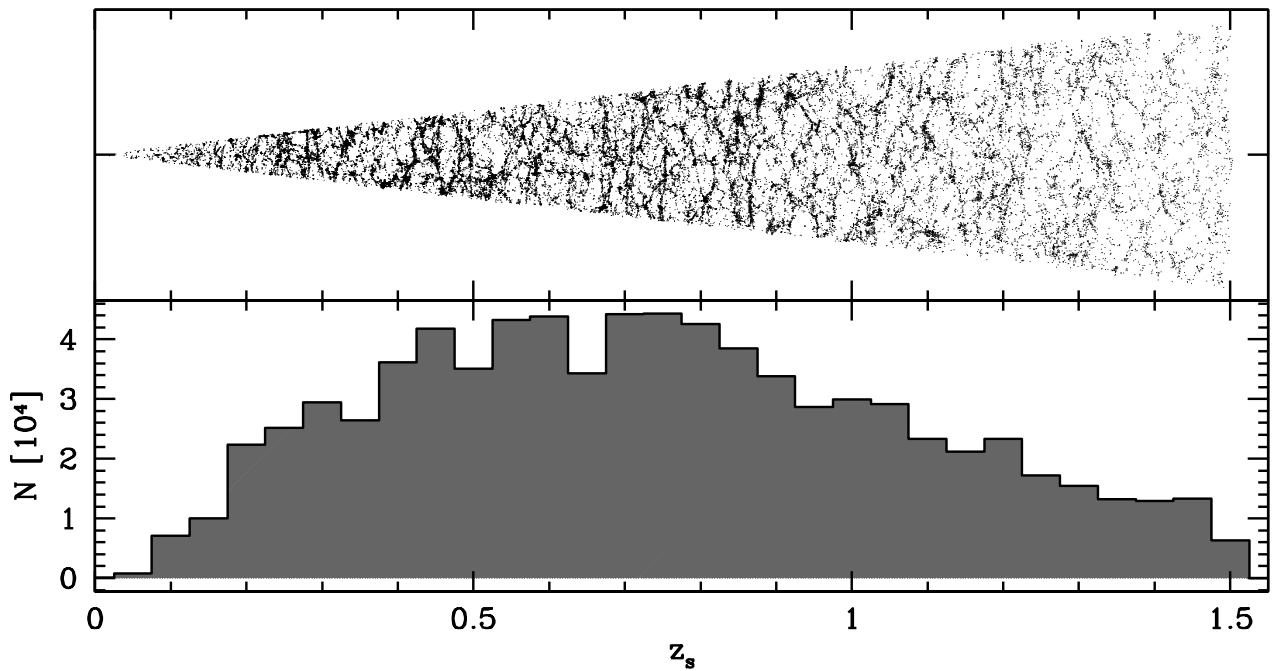}
\caption{\emph{Upper panel}: A pie plot projection showing, in a thin slice, the distribution 
of the mock galaxies as a function of redshift. 
\emph{Lower panel}: Redshift distribution of galaxies with $i_{SDSS}\leq 23$ in the selected light-cone 
with an angular coverage of 17.6 $\rm{deg}^2$. The maximum redshift of the sample is $\sim 1.5$.
}
\label{fig2}
\end{center}
\end{figure*}
We present mock observations of the artificial Universe constructed from 
the Millennium Simulation, by positioning a virtual observer at zero 
redshift and finding those galaxies which lie on the observer's backward light-cone.  
In order to do this, we  build a mock sample of galaxies within an octant 
(solid angle= $\pi/2$ sr), made up of shells taken from different snapshots 
corresponding to the epoch of the lookback time at their corresponding distance. This 
method is commonly used to construct mock galaxy catalogues and it takes 
into account gravitational evolution as well as the evolution of the 
astrophysical properties \citep{diaz02,blaizot05,kitz07,henri11,wang12}. 
We use the last 27 snapshots, which reach a maximum redshift of z=1.5.  
Given that the simulation box is only 500 $h^{-1} \ Mpc$ on a side, in order to 
reach a greater distance, it is necessary to use the periodicity of the simulation
box and build a ``super-box'', which is by construction several simulations put
together side by side.
The cosmological redshift (or redshift in real space) is obtained from the 
comoving distance of the galaxies in the super-box by using 
$r(z_c)=\int_0^{z_c} cdz'/H(z')$, where $r$ is the comoving distance and 
$H(z)=H_0\sqrt{\Omega_m(1+z)^3+\Omega_{\Lambda}}$. 

To mimic the observations, we introduce the distorted or spectroscopic redshift, 
$z_s$, by considering the peculiar velocities of the galaxies in the radial direction,
therefore:
\begin{equation}
\label{zdistorted}
z_s=(1+z_c)(1+v_p/c)-1
\end{equation}
where $z_c$ is the cosmological redshift and 
$v_p= \vec{v} \cdot \vec{r} / |\vec{r}|$ is the peculiar velocity, with $\vec{r}$ the comoving coordinate 
within the super-box (see \citealt{peacockbook99,mobook10}). 

Given that the method to construct the light-cone uses shells at different 
snapshots, it introduces differences when compared with the observed Universe:
\begin{enumerate} 
\item The first problem arises because all galaxies at a given shell have 
the same evolutionary stage corresponding to the output simulation time. 
Therefore, the mock galaxies show a discrete magnitude evolution which turns 
out to be more abrupt at higher redshifts (since the snapshots are spaced 
logarithmically with time). However, observationally, the properties of the 
galaxies vary continuously with redshift. This issue introduces a bias in the 
galaxy density distribution of the resulting mock catalogue. Also, the clustering of 
galaxies changes from snapshot to snapshot due to their proper movements: 
the larger the time-spacing between subsequent snapshots, 
the larger the variation in the structures. 
\item The second problem arises because at the edges of the imaginary shells, 
galaxies come from two different evolutionary stages. Just considering the 
movement in the simulation box, if the spacing among outputs is too large, 
the positions of the galaxies could have changed dramatically from one 
output to the next one, making a galaxy being observed either twice or 
not at all, depending on the direction of its motion (see Fig.~\ref{fig1}).
\end{enumerate}
To deal with these issues we introduced the following corrections during the 
mock construction procedure:
\begin{enumerate}
\item Positions and velocities are interpolated between the outputs in the $i$
 and $(i+1)$ shells,  according to their distance to the shell edges. 
We recompute the rest-frame absolute magnitudes $M_i$ of the galaxies within
a given shell at cosmic time, $t_i$, by interpolating linearly between the values 
corresponding to the current shell and the previous snapshot at $t_{i+1}$ 
(early time), but using the previously interpolated galaxy position 
inside the $i^{th}$ shell. It has been argued in previous works that using 
interpolated positions and velocities could produce dynamically
incorrect velocities and could diffuse structures \citep{kitz07}.
In appendix~\ref{apendA} we show that using a mock catalogue with interpolated galaxy
positions and velocities do not introduce any particular bias in the results that we have obtained
in this work.
\item We considered two possible cases. First, the repeated galaxies case, where
galaxies near the low redshift side of the $(i+1)^{th}$ shell are moving towards lower redshifts 
(top right panel of Fig.~\ref{fig1}) also appear in the $i^{th}$ shell (top left panel
of Fig.~\ref{fig1}). 
Second, the missing galaxies case, where galaxies close to the low redshift side of the 
$(i+1)^{th}$ shell, below the boundary, are moving towards higher redshifts (bottom right panel 
of Fig.~\ref{fig1}), and do not appear in the $i^{th}$ shell either (bottom left panel of the 
Fig.~\ref{fig1}). In the first case we just discarded the galaxy positioned at 
the $i^{th}$ shell, since it will appear at the consecutive shell. In the 
second case, we reassigned the position of the galaxy in the $i^{th}$ shell
with the interpolated position of the galaxy in the $(i+1)^{th}$ shell.
\end{enumerate}

As previously stated, in order to reach the desired depth of 
the catalogue we have filled the space with 
a required number of replications of the fundamental volume, 
leading us to obvious artefacts if 
the simulation is viewed along one of its preferred axes. 
Although we can not avoid this behaviour in the octant light-cone, 
we could minimise this kaleidoscopic 
effect in a smaller light-cone by orienting the survey field 
appropriately following the procedure 
described by \cite{kitz07}. According to that work, if we select an observational field 
defined by the lines-of-sight to the four points with Cartesian coordinates given by  
$( (n\pm0.5/m) L_{\rm box} , (m\pm0.5/n) L_{\rm box} , 
nm L_{\rm box} )$ where $L_{\rm box}$ is the side of the cube, and $n$ and $m$ are arbitrary numbers,
we obtain a near rectangular light-cone 
survey of angular size $1/m^2n$ x $1/n^2m$ sr with the first duplicate point at comoving distance 
$s(z_{clean})\sim mnL_{\rm box}$. In this way, we select the parameters in order to obtain a light-cone 
with a solid angle of $17.6 \ deg^2$ and without repetitions out to $z\sim 1.02$.

The volume limited sample with absolute magnitudes brighter than $-16.4$ contained in the selected 
light-cone comprises 6,756,097 galaxies up to $z=1.5$. 
Finally, we compute the observer-frame galaxy apparent magnitudes from the 
publicly available rest-frame absolute magnitudes provided by the semi-analytic model:
$m=M+25+5\log\left(s(1+z_s)\right) - k_{corr}(z_s)$,
where $s$ is the comoving distance computed from the spectroscopic redshift. 
The $k$-corrections are obtained as a byproduct of the method that 
computes the photometric redshifts (see Sect.~\ref{bpz_method}). 
We set an observer-frame apparent magnitude limit of $i_{lim}=23$. 

The final spectroscopic mock catalogue (\emph{sp-mock}) comprises 793,559 
galaxies with a median redshift of $0.72$ within a solid angle of 17.6 $deg^2$.
In Fig.~\ref{fig2} we show an illustration of the galaxy distribution 
as a function of redshift (\emph{upper panel}) 
and  the redshift distribution of galaxies with $i_{SDSS}\leq 23$ in the 
selected light-cone (\emph{lower panel}).

\begin{figure}
\begin{center}
\includegraphics[width=60mm]{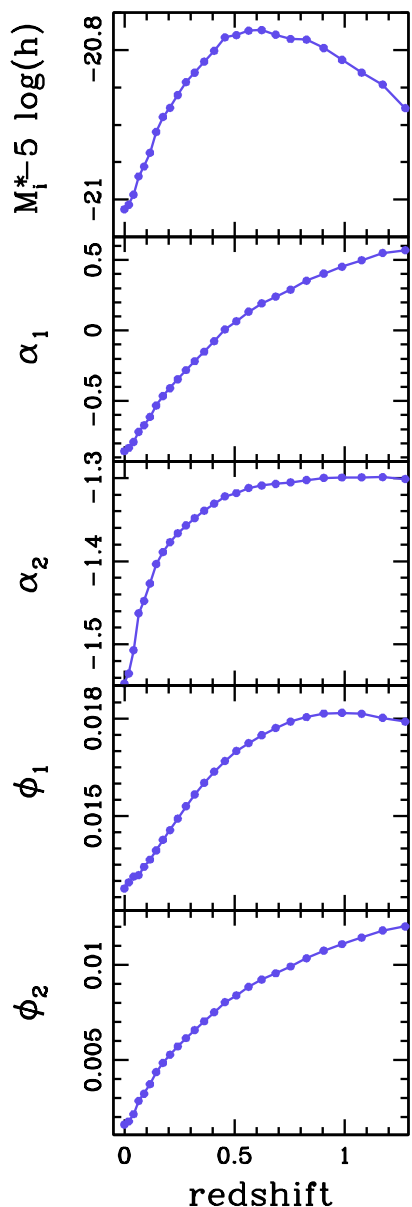}
\caption{The best fitting parameters of a double Schechter Luminosity function in the $i_{SDSS}$ band 
as a function of redshift. These parameters are obtained by fitting the luminosity distributions of the 
semi-analytic galaxies \citep{guo11} down to an absolute magnitude of $-16.4$ 
at different snapshots of the Millennium Simulation.
}
\label{LF_params}
\end{center}
\end{figure}
\subsection{Photometric redshift assignment}
\label{bpz_method}

We assigned photometric redshifts to the mock catalogue previously built. 
In order to do this, we first obtained spectral types from the original rest-frame photometry 
and spectroscopic redshifts by running the Bayesian Photometric Redshift package (BPZ, \cite{BPZ}) with 
the ONLY\_TYPE yes option. Then, we transformed the given photometry in the mock catalogue 
to the photometry of the J-PAS.
This transformation uses the filter curve response and the spectral types obtained.
Finally, we ran again BPZ on this new photometry obtaining the photometric redshift associated 
with the new photometry. 
As a byproduct of this method, we compute the observer-frame apparent magnitudes of the mock galaxies
(and therefore, their corresponding k-corrections). All the details can be found in Ascaso et. al (2013, in prep.) 

\section{The Friends-of-Friends algorithm and the tuning of the linking length parameters}

The Friends-of-Friends algorithm was initially developed to identify galaxy systems in redshift space 
considering a flux limited catalogue \citep{huchra82}. Since then, several adaptations of 
this percolation algorithm have been used \citep{merchan02,eke04,knobel09}, or modified 
to identify halos in 3-D from simulations \citep{davis85} - for a compilation of algorithms see 
\cite{knebe11}) - or identifying groups through photometric redshifts 
\citep{botzler04,li08,liu08}. 

The FoF algorithm links galaxies that share common neighbours (friends).
It starts looking for the "friends" of an initial galaxy that 
have separations lower than a given threshold. Groups are defined as sets of galaxies
that are connected by one or more friendship relations, i.e., friends of friends.
For each galaxy not assigned to a group, the algorithm searches around it for 
companions with projected separation from the first galaxy: 
$$D_{12} = 2 \, R_{12} \tan{\frac{\Theta_{12}}{2}} \le D_l$$ 
and 
$$V_{12} = | V_1-V_2| \le V_l$$

where $\Theta$ is the angular separation among a pair of galaxies, $V_1$  and $V_2$ 
refer to their radial velocities (or redshifts), 
and $R_{12}=(R1+R2)/2$ is the mean of their comoving distances.
All friends of a galaxy are added to the list of group members. The surroundings of 
each friend are then examined. This process is repeated until no further neighbours 
are found.

When working with observational samples, there are two main characteristics inherent 
to the observations that make the group finding difficult. One of them is the flux 
limit of the catalogue, and the other is the redshift space distortion. In order to 
adopt the best linking length parameters, $D_l$ and $V_l$,
both issues must be treated separately.

\begin{table*}
\begin{center}
\tabcolsep 1pt
\caption{Groups identified in different mock galaxy samples \label{grupos}}
\begin{tabular}{llllrrr}
\hline
Sample      & flux limit & space             & Linking lengths  & Total number of & Groups with   & Groups with \\ 
          & $i_{SDSS}$&               &                  & gals in groups  & $4\le N < 10$ & $N \ge 10$  \\
\hline
\emph{reference}   &  &  real           & $D_l=D_0(z) \hspace{1 cm} \ \  V_l=D_l \, H(z)$          & 1,825,303 & 159,258 & 41,774 \\
\emph{restricted-reference}   & 23 & real &                                                       &  120,256 & 11,648 &  2,699 \\
\emph{flux limited}-LF variable & 23 & real & $D_l=D_0(z) \, R_s(z) \ \ \ \ \ V_l=D_l \, H(z)$         &   138,675 &  14,317 &  2,980 \\
\emph{flux limited}-LF fixed    & 23 & real & $D_l=D_0(z) \, R_s(0) \ \ \ \ \ V_l=D_l \, H(z)$         &   159,484 &  16,641 &  3,414 \\
Redshift                 &  & sp-redshift & $D_l=D_0(z) \hspace{1cm} \ \  V_l=130$                   & 1,287,097 & 160,145 & 23,572 \\ 
                         &  & sp-redshift & $D_l=D_0(z) \hspace{1cm} \ \  V_l=130 (1+z) $            & 2,133,189 & 203,975 & 46,557 \\ 
                         &  & sp-redshift & $D_l=D_0(z) \hspace{1cm} \ \  V_l=70 $                   &   629,841 & 98,537 & 8,383 \\ 
                         &  & sp-redshift & $D_l=D_0(z) \hspace{1cm} \ \  V_l=70 (1+z) $             & 1,394,091 & 170,918 & 26,372 \\
sp-mock catalogue        & 23 & sp-redshift & $D_l=D_0(z) \, R_s(z) \ \ \ \ \ V_l=130 (1+z) \, R_s(z)$ &   172,367 &  19,780 &  3,403 \\
\hline
\end{tabular}
\end{center}
\end{table*}

\subsection{Reference sample: volume limited sample in real space}
\label{reference-samples}
\begin{figure}  
\begin{center}
\includegraphics[width=\hsize]{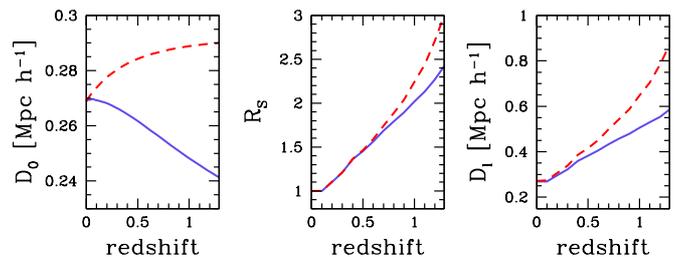}
\caption{The variation of the linking length parameters as a function of redshift.
The \emph{left panel} shows the transversely linking length for volume limited samples, $D_0$, 
the \emph{middle panel} shows the scale factor $R_s$, while the \emph{right panel} shows the 
transverse linking length for flux-limited samples, $D_l$ (see equations in Sect.~\ref{flrs}).  
The \emph{blue solid lines} show the parameters using a LF that varies with redshift
(see Fig.~\ref{LF_params}) while \emph{red dashed lines} show the parameters when a 
fixed LF at redshift close to zero is adopted. 
}
\label{params}
\end{center}
\end{figure}

We define a sample of galaxies without the two mentioned issues, i.e., we created a 
volume limited sample of galaxies in real space. This sample is complete down to 
absolute magnitude $M_{i_{SDSS}}=-16.4$. Avoiding the observational constraints, the 
identification of groups in this sample can be performed straightforwardly. The 
linking length parameters are defined as follows:
\[ D_l=D_0 \ \ {\rm and} \ \ V_l=D_l \, H(z) \] 
where $H(z)$ is the Hubble constant as a function of redshift and $D_0$ takes into 
account the overdensity of virialised structures in the Universe at a given time:
\begin{equation}
\label{d0}
D_0(z)= \left[ \frac{4\pi}{3} \left( \frac{\delta \rho}{\rho} (z) + 1 \right) 
\int^{M_{lim}}_{-\infty} \phi(z,M) dM \right] ^{-1/3}
\end{equation}
where $\phi(z,M)$ is the luminosity function, and $\frac{\delta \rho}{\rho}(z)$
is the contour overdensity contrast. 
Similar to other authors in previous works (see for instance, \citealt{snaith11}), 
in order to model the $\frac{\delta \rho}{\rho}(z)$, we assume that galaxies
are unbiased mass tracers.
Analysing the mass function of halos identified with FoF algorithms, \cite{courtin11} found 
deviations from universality in the mass function due to the use of halo parameters not adjusted 
for different virialisation overdensities in different cosmologies and redshifts. 
\cite{more11} showed that the boundary of FoF halos does not correspond to a single 
local overdensity, but rather to a range of overdensities, and that the enclosed overdensities 
of the FoF halos are significantly larger than commonly thought. \cite{courtin11} showed 
that deviations from universality are not random but are correlated with the nonlinear 
virialisation overdensity, $\Delta_{vir}$, expected from the spherical collapse model for a given 
cosmology and redshift. In particular, they showed that the linking length required to minimise 
deviations of the FoF mass function from universal form for a given cosmology and redshift is 
correlated with the corresponding $\Delta_{vir}$ as:
\begin{equation}
\label{rho}
\frac{\delta \rho}{\rho}(z) = b^{-3}(z) = b^{-3}_{0} 
\left(0.24 \frac{\Delta_{vir}(z)}{178} + 0.68  \right)
\end{equation}
where $b_{0}$ is the linking length parameter commonly used for identifying dark matter 
halos and is set to a value of $0.2$. 
From \cite{weinberg03}, the enclosed overdensity of virialised halos is
\[ \Delta_{vir} (z) = 18 \pi^2 \left[ 1 + 0.399\left( \frac{1}
{\Omega_m(z)} - 1 \right)^{0.941}\right] \]
with
$ \left( \frac{1}{\Omega_m(z)} - 1 \right) = \left( \frac{1}{\Omega_0}-1 \right) 
(1+z)^{-3} $.
For a Universe with cosmological parameters (0.3, 0.7), the last equation leads to 
the known value of an enclosed overdensity of virialised halos at z=0 of $\sim 330$. 
It is worth reminding that for the Millennium simulation the cosmological parameters 
are (0.25, 0.75) which implies that the virialised overdensity at z=0 is $\sim 376$. 

Even though we are adopting a redshift dependent contour overdensity contrast for
our algorithm, it is worth noting that, for the cosmology of the Millennium Simulation, 
the empirical relation produces a variation
of $b(z)$ of only $\sim 8\%$ in the whole redshift range under study.
On the other hand, in appendix~\ref{apendB} we introduce a variation in the 
eq.~\ref{rho} in order to investigate the effect in our results of using a higher contour 
overdensity contrast, as expected from the analyses of galaxy group density profiles. 

Before applying the identifier, it is necessary to compute the luminosity function 
of the galaxies in the catalogue. To this end, we made use of the information from 
the semi-analytic model, and computed the LF for every snapshot of the simulation. 
Then, we fitted double-Schechter functions to the distributions of rest-frame $i_{SDSS}$ 
absolute magnitudes: $$\phi(L)=\frac{1}{L^*} exp \left(-\frac{L}{L^*}\right) \, 
\left[ \phi_1 \left(\frac{ L}{ L^*} \right)^{\alpha_1} + 
\phi_2 \left(\frac{ L}{ L^*} \right)^{\alpha_2} \right]$$
The best fitting parameters are shown in 
Fig.~\ref{LF_params} as a function of the redshift.

The variation of $D_0$ used in this section as a function of redshift can be seen as 
the \emph{solid line} in the left panel of Fig.~\ref{params}.

This algorithm produces a sample of $201,032$ groups with 4 or more galaxy members,
within a solid angle of $17.6 \ \rm{deg}^2$ up to redshift $1.5$ (see Table~\ref{grupos}). 
These groups constitute the \emph{reference} sample that will be used for testing the 
algorithm against as we introduce the observational constraints in the mock catalogue.

It is also worth selecting from the \emph{reference} groups those 
that have 4 or more members with observer-frame magnitude $i_{SDSS}$ brighter than 23, i.e., 
those groups that could be identified in the flux limited catalogue. 
We will refer to this subsample of reference groups as \emph{restricted-reference} group sample, 
which comprises $14,347$ groups (see Table~\ref{grupos}).

\subsection{Flux limited sample in real space}
\label{flrs}
\begin{figure*}
\centering
{\includegraphics[scale=0.4,clip=]{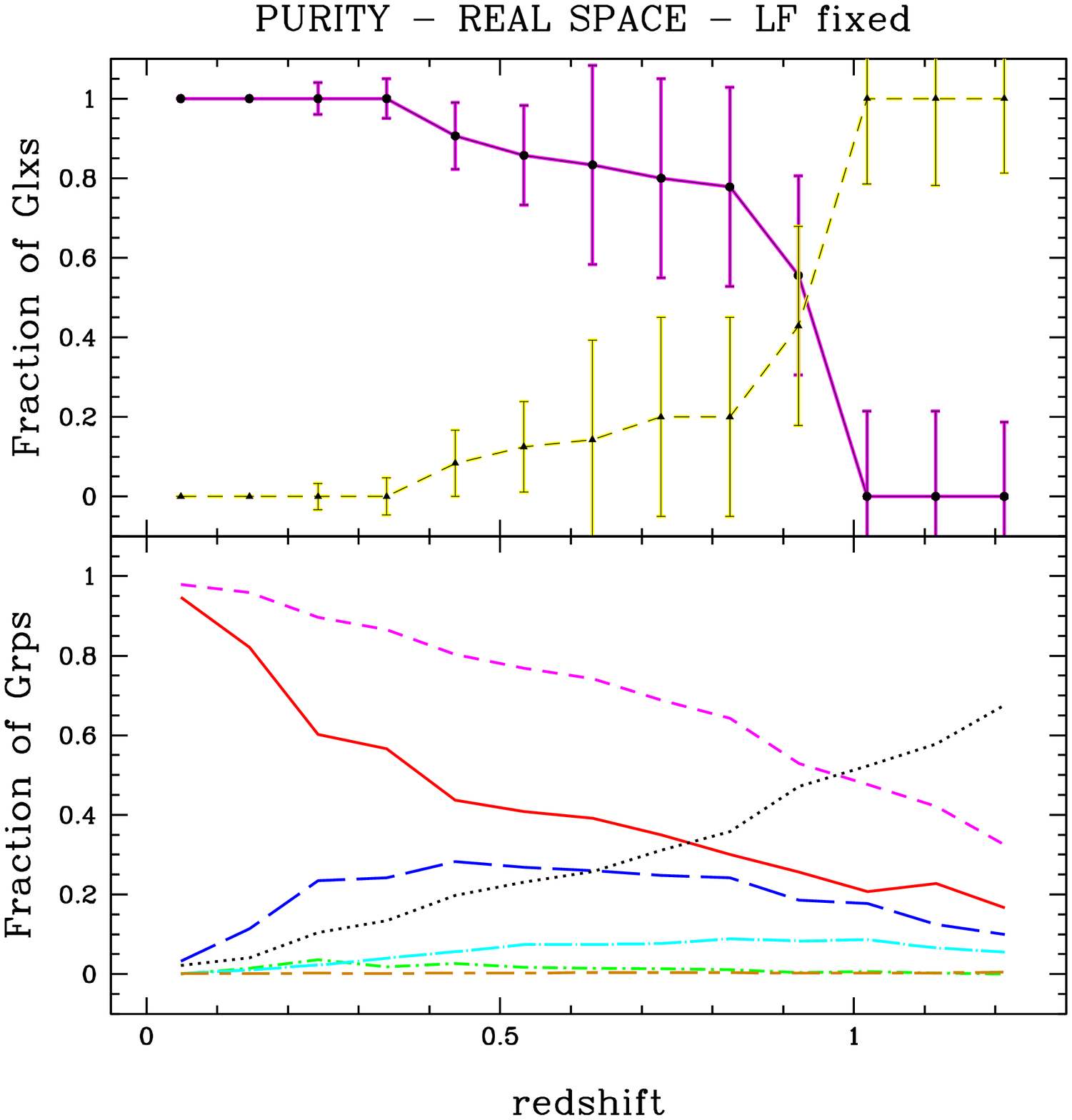}
\includegraphics[scale=0.4,clip=]{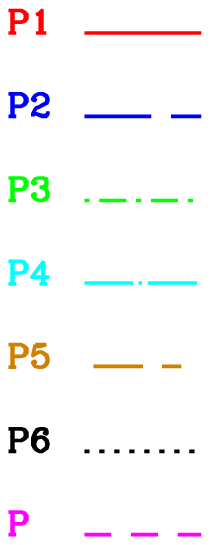}
\includegraphics[scale=0.4,clip=]{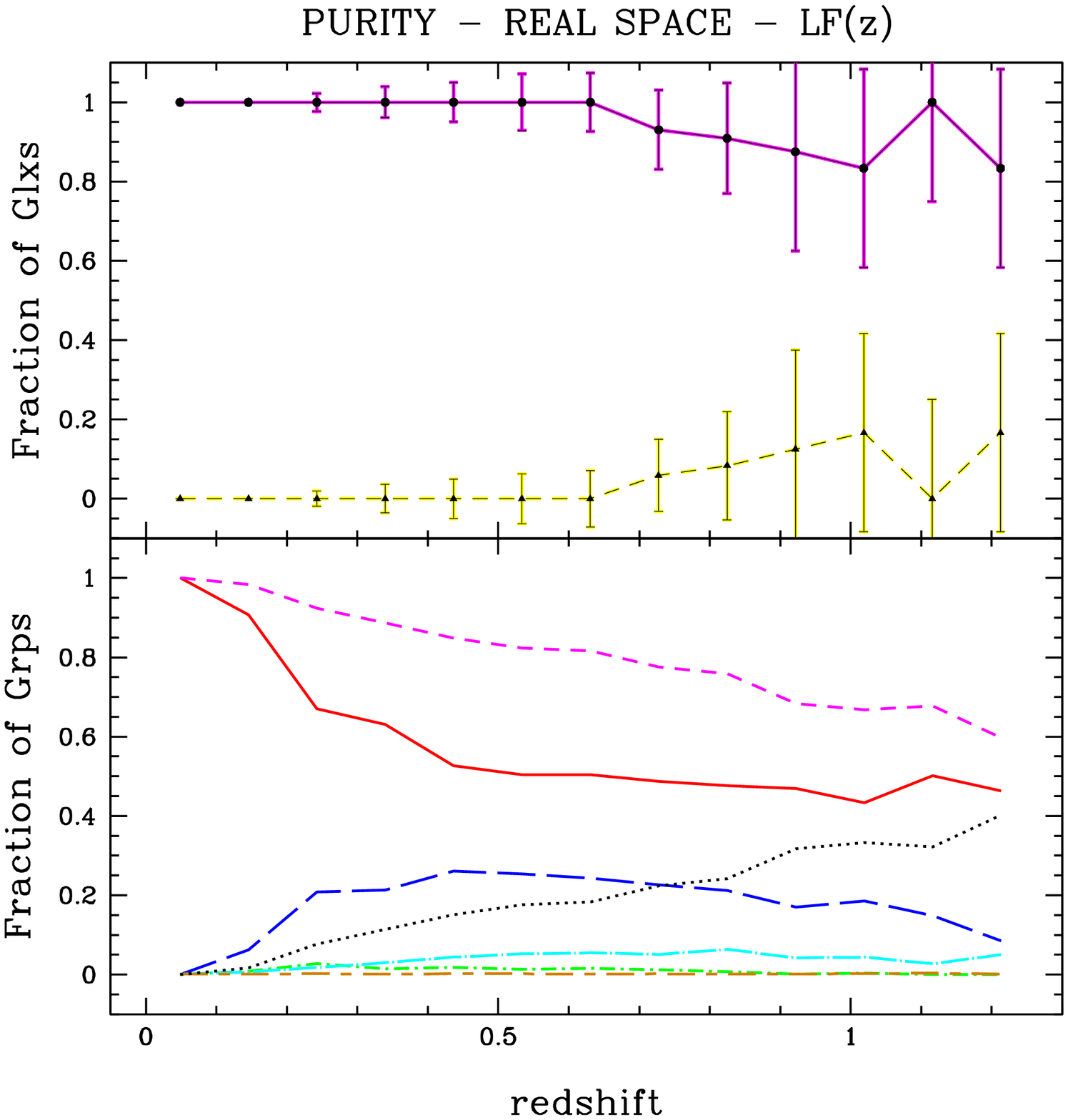}}
{\includegraphics[scale=0.4,clip=]{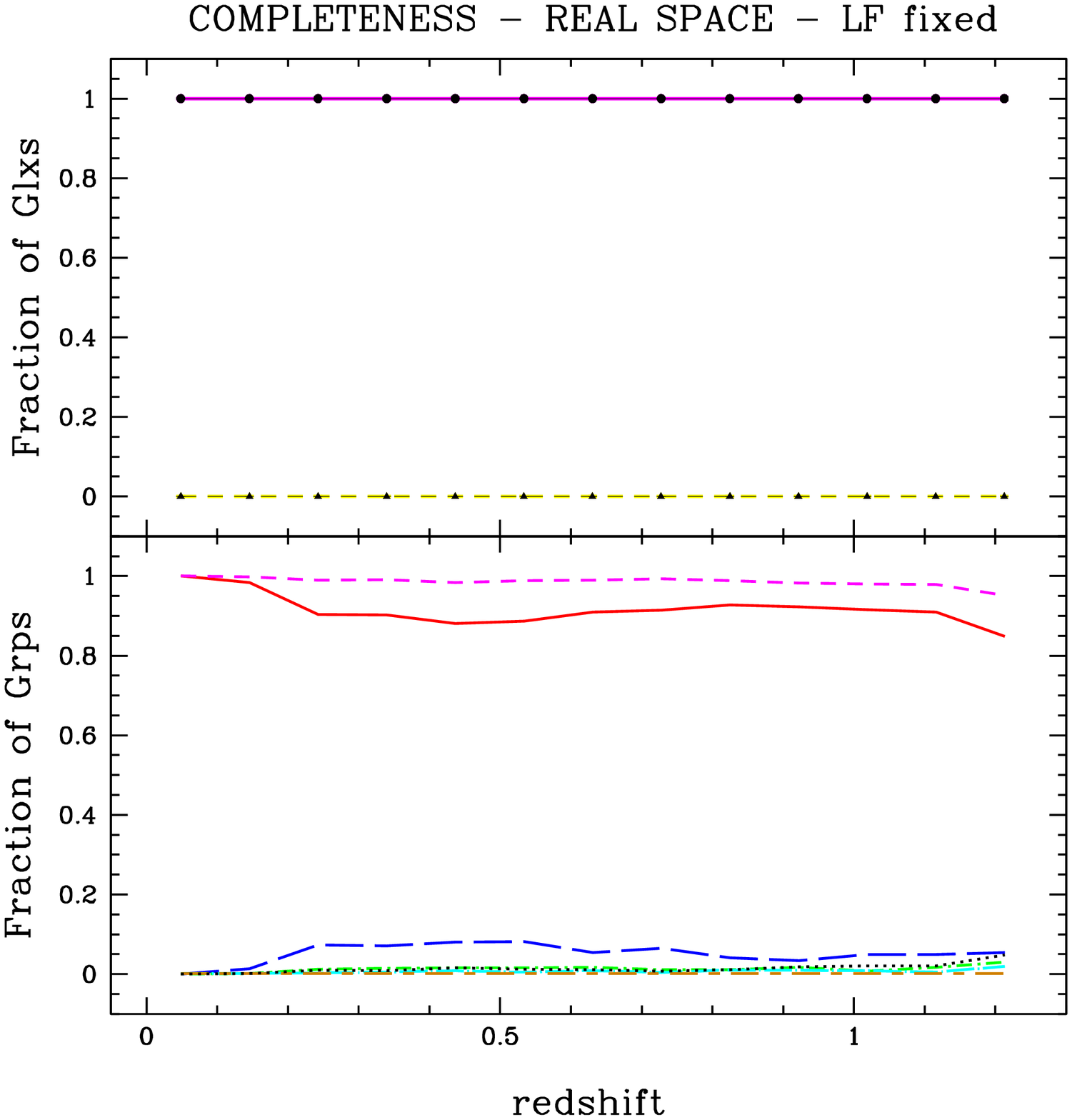}
\includegraphics[scale=0.4,clip=]{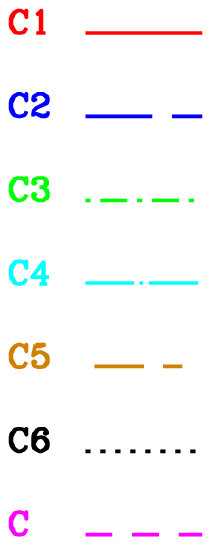}
\includegraphics[scale=0.4,clip=]{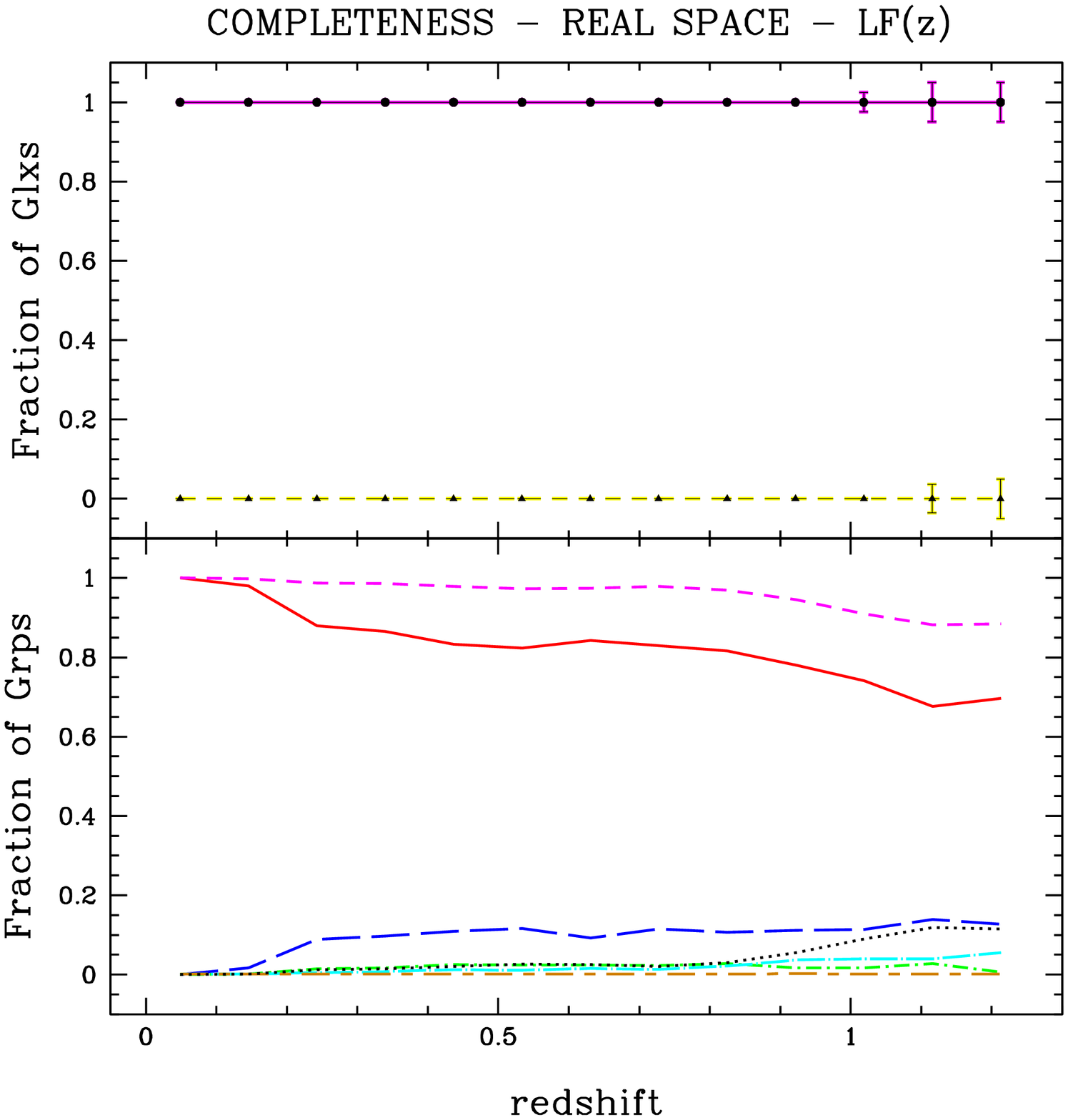}}
\caption{Purity and completeness as a function of redshift for groups identified
in a \emph{flux limited} sample in real space. \emph{Left boxes} show the purity (\emph{upper}) and
completeness (\emph{lower}) when a fixed LF is used in the linking length parameters,
while right boxes show the trends when a LF varying with redshift is 
used to compute the linking length parameters. 
In the upper boxes (\emph{purity}), the top panels show the fraction of identified galaxies (Glxs)
 associated with
the group with the largest matching rate in the corresponding control sample (\emph{solid lines}), 
and the fraction of galaxies that are classified as interlopers (\emph{dashed lines}). 
In the lower boxes (\emph{completeness}), the top panels show the fraction of galaxies in the control sample 
that are associated with the group with the largest matching rate in the identified sample 
(\emph{solid lines}), 
and the fraction of galaxies that are classified as missing galaxies (\emph{dashed lines}). 
The \emph{bottom panels} in each plot show the trends observed for the fraction of groups (Grps) within 
the six categories of \emph{purity} or \emph{completeness} (see text for description). The \emph{magenta short dashed lines} correspond to the complement of $P6$ and $C6$.  
\label{purity_flux}}
\end{figure*}

We first tested the algorithm against a flux limited sample. Now, both linking lengths 
have to take into account the flux limit of the catalogue, so besides being related 
to the overdensity contrast they have to include the variation of the sampling of the 
luminosity function produced by the different distances of the groups to the observers, 
which is introduced, following \cite{huchra82}, 
by the scale factor $R_s$\footnote{We kept the notation introduced by \cite{huchra82} 
although the parameters in this work depend also on the redshifts.}:
\[ D_l=D_0 \, R_s \ \ {\rm and} \ \ V_l= D_l \, H(z) \]
with
\begin{equation}
\label{rs}
R_s (z) = \left[ \frac{\int^{M_{12}}_{-\infty} \phi(z,M)dM}{\int^{M_{lim}}_
{-\infty} \phi(z,M)dM} \right]^{-1/3}
\end{equation}
where 
$M_{lim}=-16.4$,
and $M_{12}=i_{lim} - 25 - 5 \log{ ( {d_L}_{12} )}$, with $ {d_L}_{12}$ the 
mean luminosity distance for the galaxy pair.  

\begin{table*}
\begin{center}
\tabcolsep 5pt
\caption{Total percentages of purity and completeness of groups (with $z= 0 -1.2$) identified in different mock galaxy samples \label{percentage}}
\begin{tabular}{ccccccccccc}
\hline
Class & & \multicolumn{2}{c}{Flux limited} & &  \multicolumn{4}{c}{Redshift} & & Sp-mock \\ 
      & & LF fixed & LF variable  & &   $V_0=130$ &  $V_0=130(1+z)$ & $V_0=70$  & $V_0=70(1+z)$ & & $V_0=130(1+z)$\\
\cline{1-1} \cline{3-4} \cline{6-9} \cline{11-11}
P1   & & 39 & 54 & & 42 & 35 & 49 & 42 & & 22 \\
P2   & & 23 & 22 & & 21 & 21 & 20 & 21 & & 19 \\   
P3   & & \ 2 & \ 1 & & \ 6 & \ 5 & \ 8 & \ 6 & & \ 2 \\
P4   & & \ 6 & \ 4 & & 12 & 11 & 11 & 12 & & 13 \\
P5   & & \ 0 & \ 0 & & \ 2 & \ 1 & \ 1 & \ 1 & & \ 1 \\
P6   & & 30 & 19   & & 17 & 27 & 11 & 18 & & 43 \\
\hline
C1   & & 91 &  84 & & 14 & 48 & \ 3 & 17 & & 44 \\
C2   & & \ 6 & 10 & & 16 & 25 & \ 5 & 18 & & 25 \\
C3   & & \ 1 & \ 2 & & \ 6 & \ 8 & \ 2 & \ 7 & & \ 8 \\
C4   & & \ 1 & \ 1 & & 21 & \ 8 & 19 & 21 & & 10 \\
C5   & & \ 0 & \ 0 & & \ 6 & \ 1 & \ 7 & \ 5 & & \ 1 \\
C6   & & \ 1 & \ 3 & & 37 & 10 & 64 & 32 & & 12 \\
\hline
\end{tabular}
\end{center}
\end{table*}

Usually, for low redshift samples, the luminosity function of galaxies included in 
the $R_s$ factor is computed for the whole sample, and it is assumed that there is no 
evolution in the luminosities up to the maximum depth of the catalogue. Since we intend
to reach higher redshift groups, it is worth introducing the evolution 
of the luminosities of the catalogued galaxies. Therefore, it is important to compute 
the luminosity function of the galaxies in bins of redshifts, as we did in the previous 
section, to account for the variation of the density of galaxies as well as their 
internal luminosity evolution. However, in this section we will also select a sample of groups 
without using the luminosity evolution of galaxies, i.e, by using a fixed luminosity function 
determined at redshift zero to assess the importance that it could have in the 
resulting sample. In Fig.~\ref{params}, the variation of $D_0$, $R_s$ and the 
linking length $D_l$ are shown as a function of redshift. 
\emph{Solid lines} correspond to the values obtained from a LF that varies with redshift, 
while \emph{dashed lines} correspond to a fixed luminosity function. 

We use an observer-frame apparent magnitude $i_{SDSS}^{lim}=23$ to limit our mock galaxies. 
The number of groups with 4 or more members identified with a fixed LF is $20,055$, 
while when varying the LF with redshift, it is $17,297$ (see Table~\ref{grupos}).

In order to compare the sample of groups identified in this flux limited catalogue to 
the reference sample, we use the \emph{restricted-reference} sample 
to analyse the purity and completeness of the \emph{flux-limited} groups.

We define \emph{purity} and \emph{completeness} based on a member-to-member comparison. 
As \emph{purity}, we consider the fraction of members in the flux-limited groups that belongs 
to any \emph{restricted-reference} group, i.e., we want to quantify how good the identified groups are. 
As \emph{completeness} we consider the fraction of members in the \emph{restricted-reference} 
groups that are part of the flux-limited groups, 
this quantity intends to indicate the fraction of the true groups that we are able to identify. 

Regarding the \emph{purity} of the flux-limited sample, in the upper panels of 
Fig.~\ref{purity_flux} we show the fraction of galaxies belonging
to a flux-limited group that are associated to one \emph{restricted-reference} group which possesses
the largest matching rate (\emph{solid lines}), and the fraction of flux-limited group galaxy 
members that are not associated to any \emph{restricted-reference} group 
(interlopers, \emph{dashed lines}), both as a function of their real-space redshifts. 
The left boxes correspond to the flux-limited sample identified with a 
fixed LF, while the right boxes correspond to the sample identified with LF variable.
From these plots, it is clear that the effect of assuming no-evolution in the luminosities 
leads to a more contaminated sample towards higher redshifts.
It can be seen that, when considering evolution in the luminosity function, 
the \emph{purity} of our flux-limited groups is high, or in other words,
the fraction of interlopers is really low (less than 20\%).

However, quantifying the fraction of member galaxies in a flux-limited group that belong to some
\emph{restricted-reference} group is not enough to understand the real nature of the identified groups.
For instance, one single flux-limited group could be formed by members that 
originally belonged to more than one \emph{restricted-reference} group.
In order to disentangle the different galaxy contributions to a given galaxy group,
six group categories could be defined when comparing two samples of groups: ${\cal A}$ and ${\cal B}$. 
\begin{enumerate}
\item $P1$ (perfect match): Groups in sample ${\cal A}$ having 100\% of their members associated 
with only one 
group in the control sample ${\cal B}$ (\emph{red solid lines})
\item $P2$ (quasi-perfect match): Groups in sample ${\cal A}$ having between 70\% and 100\% of their 
members associated with only one group in the control sample ${\cal B}$, and the remaining galaxies are interlopers 
(0\%$<$interlopers$<$30\%) (\emph{blue long dashed lines})
\item $P3$ (merging): Groups in sample ${\cal A}$ having between 70\% and 100\% (inclusive) of their 
members associated with more than one group in the control sample ${\cal B}$. This category may accept 
interlopers (0\%$\le$ interlopers $<$ 30\%) (\emph{green dot short dashed lines})
\item $P4$ (group+interlopers): Groups in sample ${\cal A}$ having less than 70\% of their members belonging 
to only one group in the control sample ${\cal B}$. The remaining members are interlopers (interlopers$>30\%$)
(\emph{cyan dot long dashed lines})
\item $P5$ (merging+interlopers): Groups in sample ${\cal A}$ having less than 70\% of their members 
belonging to more than one group in the control sample ${\cal B}$, the remaining galaxies are interlopers 
(\emph{brown short dash-long dashed lines})
\item $P6$ (false): Groups in sample ${\cal A}$ having 100\% of their members not belonging to 
any group in the control sample ${\cal B}$ (100\% interlopers) (\emph{black dotted lines})
\end{enumerate}

In this case, to examine the \emph{purity} of the \emph{flux-limited} groups, they are split 
into the six categories defined above taking the sample ${\cal A}$ as the \emph{flux-limited} 
sample, while the control sample ${\cal B}$ is the \emph{restricted-reference} sample.

The fractions of \emph{flux-limited} groups within each category of purity per redshift bin are shown in 
the bottom panels of the upper boxes of Fig.~\ref{purity_flux}. 
The ``perfect match'' between \emph{flux-limited} and \emph{restricted-reference} groups are 
those in the $P1$, in which all the group members of the \emph{flux-limited} sample belong to a 
unique \emph{restricted-reference} group (still, the \emph{restricted-reference} group might 
have more extra members).
As expected, the higher the redshifts, the lower the fraction of perfectly matched groups. 
Even though this behaviour is common for both identifications, the $P1$ sample when using a variable
LF has a higher percentage of groups along the whole redshift range than the corresponding values for the
fixed LF. The $P2$ sample includes ``quasi''-perfectly matched groups. The fraction of these groups is similar 
in both identifications. 

The \emph{green dot short dashed} lines ($P3$) involve \emph{flux-limited} groups that are the 
result of merging true groups plus few interlopers. 
For both identifications, this category is almost nonexistent.

The $P4$ and $P5$ contain part of real groups, but also the interlopers are an important 
fraction of the galaxies in these groups. In both identification they sum up less than $\sim 10\%$ in the whole 
redshift range.  

The least desired category is $P6$, those are completely false groups. It can be seen that identifying with a 
fixed LF produces a higher percentage of false groups at higher redshifts. 
The \emph{magenta short dashed} lines are the complement of the $P6$ class, 
therefore represent all other groups except the least desirable class, $P6$, 
or in other words, groups that contain at least part of the true groups.

In Table~\ref{percentage} we quote the percentage of groups in each of these classes 
for the whole samples. 
It can be seen that the sample of \emph{flux-limited} groups obtained 
from a variable LF contains a higher percentage ($+15 \%$) of $P1$ groups, 
lower percentage ($-11 \%$) of $P6$, 
and very similar percentages of the remaining classes 
than those obtained when identifying groups with a fixed LF. 

Regarding the \emph{completeness} of the sample, the lower plots of Fig.~\ref{purity_flux} show 
the results as a function of redshifts for both samples, fixed LF (\emph{left panels}) and 
variable LF (\emph{right panels}). To define \emph{completeness}, we quantify how many of the 
\emph{restricted-reference} groups were identified in the \emph{flux-limited} sample. 
From the upper panels of the \emph{completeness} plots, it can be seen than more than $95\%$ of 
the members of the \emph{reference-sample} are included in a given group of the 
\emph{flux-limited} samples. 

Following a similar procedure as used for the \emph{purity} analysis, we split groups into six 
completeness categories: 
\begin{enumerate}
\item $C1$ (perfect match): Groups in the control sample ${\cal A}$ having 100\% of 
their members identified within only one 
group in the sample ${\cal B}$ (\emph{red solid lines})
\item $C2$ (quasi-perfect match): Groups in the control sample ${\cal A}$ having between 
70\% and 100\% of their members 
identified within only one group in sample ${\cal B}$, 
and the remaining galaxies are missing in the new
identification (0\%$<$missing$<$30\%, \emph{blue long dashed lines})
\item $C3$ (split): Groups in the control sample ${\cal A}$ having between 70\% and 100\% (inclusive) of their members 
identified within more than one group in sample ${\cal B}$. 
This category may accept missing galaxies (0\%$\le$ missing $<$ 30\%, \emph{green dot short dashed lines})
\item $C4$ (group+missing galaxies): Groups in the control sample ${\cal A}$ having less than 70\% of their 
members identified within only one group in sample ${\cal B}$. 
The remaining members are not identified in any group in the new identification
 (missing$>30\%$, \emph{cyan dot long dashed lines})
\item $C5$ (split+missing galaxies): Groups in the control sample ${\cal A}$ having less than 70\% of 
their members identified within more than one group in sample ${\cal B}$, 
the remaining galaxies are lost (\emph{brown short dash long dashed lines})
\item $C6$ (missing group): Groups in the control sample ${\cal A}$ having 100\% of their members not identified in 
any group in sample ${\cal B}$ (100\% missing galaxies, \emph{ black dotted lines})
\end{enumerate}

\begin{figure}
\centering
\includegraphics[width=\hsize]{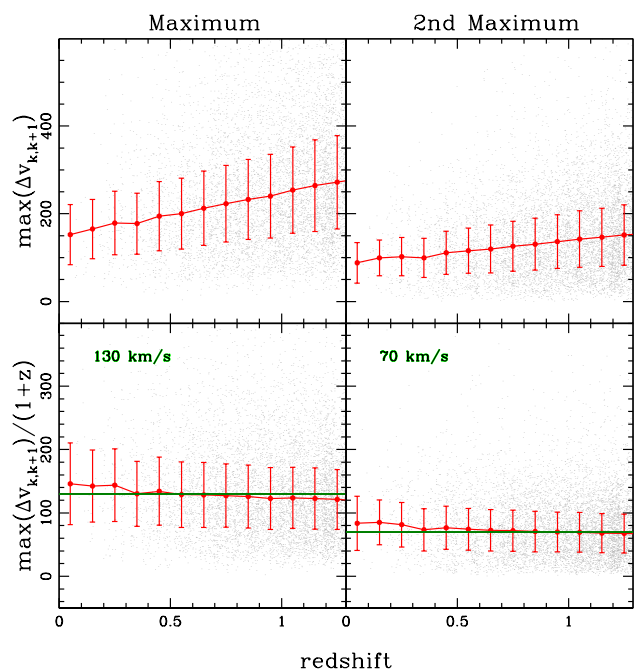}
\caption{\emph{Left upper panel}: Scatter plot of the maximum velocity difference 
of the members in the line-of-sight to their closest neighbours. \emph{Right upper panel}: 
the same as the \emph{right panel} but using the second maximum.
In the \emph{lower panels} we divided the upper panels by $(1+z)$.
\label{delta_v}}
\end{figure}

In this case, the control sample ${\cal A}$ is the \emph{restricted-reference} 
sample, while the sample ${\cal B}$ is the \emph{flux-limited} group sample.
The \emph{completeness} as a function of redshifts based on the different categories is shown in the
\emph{lower plots} of the bottom boxes of Fig.~\ref{purity_flux}.
We find that both algorithms are able to identify most of the members of the 
\emph{restricted-reference} sample, i.e., the $C1$ and $C2$ categories are dominant at all redshifts.
We observe that the
variable LF identification shows a more pronounced decay of the fractions of $C1$ groups
 to higher redshift than the observed for the fixed LF case,
however, this behaviour is almost fully compensated for an increasing fraction of $C2$ groups. 
The fraction of groups in the other categories is almost negligible, with a slightly increase of
 $C6$ groups towards higher redshifts in the variable LF identification, 
being lower than $20\%$ at the highest redshifts. 
  
In Table~\ref{percentage} we quote the total percentages of \emph{restricted-reference} groups belonging 
to each of the completeness categories. The $C1$ class is $7 \%$ lower and the $C2$ is $4\%$ higher
in the variable LF identification than in the fixed LF identification, while the $C6$ are quite similar in both .
One might be tempted to think that the identification with fixed LF produced a better 
result since the fraction of $C1$ groups in this identification is slightly higher 
and the fraction of $C6$ slightly lower than when using a variable LF. 
However, it is not worth recovering most of the true group members if the identified groups will 
be contaminated by a larger number of interlopers that could change the intrinsic properties of the groups or including many false groups.
Therefore, it is important to analyse the combination of \emph{purity} and \emph{completeness}.
The categories~$1$ and~$2$ represent the highly \emph{pure} and \emph{complete}. Analysing 
Table~\ref{percentage}, the percentages of \emph{highly complete} groups of both identifications are 
quite similar ($94\%$ vs. $97\%$), while the percentage of \emph{highly pure} groups when 
identified with LF variable is $13\%$ higher. Also, the fixed LF produces $30\%$ of false groups 
compared with $19\%$ for the variable LF. 
Therefore, using a variable LF to identify groups is the most appropriated procedure 
to recover as best as possible most of the \emph{restricted-reference} group sample. 

\subsection{Redshift distortions: volume limited sample in spectroscopic redshift space}
\label{zdistortiden}
\begin{figure*}
\centering
{\includegraphics[scale=0.40,clip=]{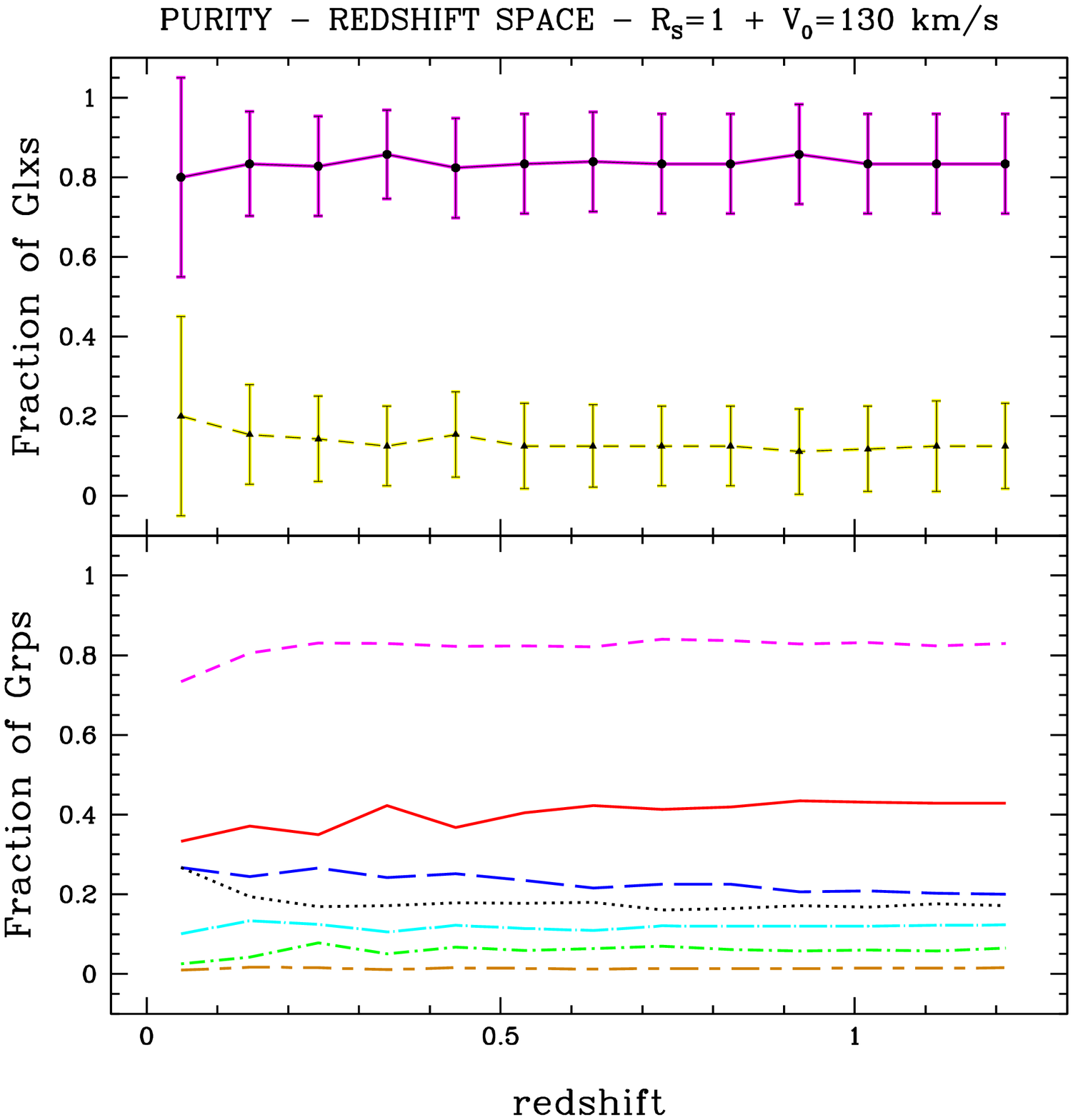}
\includegraphics[scale=0.40,clip=]{p_label.eps}
\includegraphics[scale=0.40,clip=]{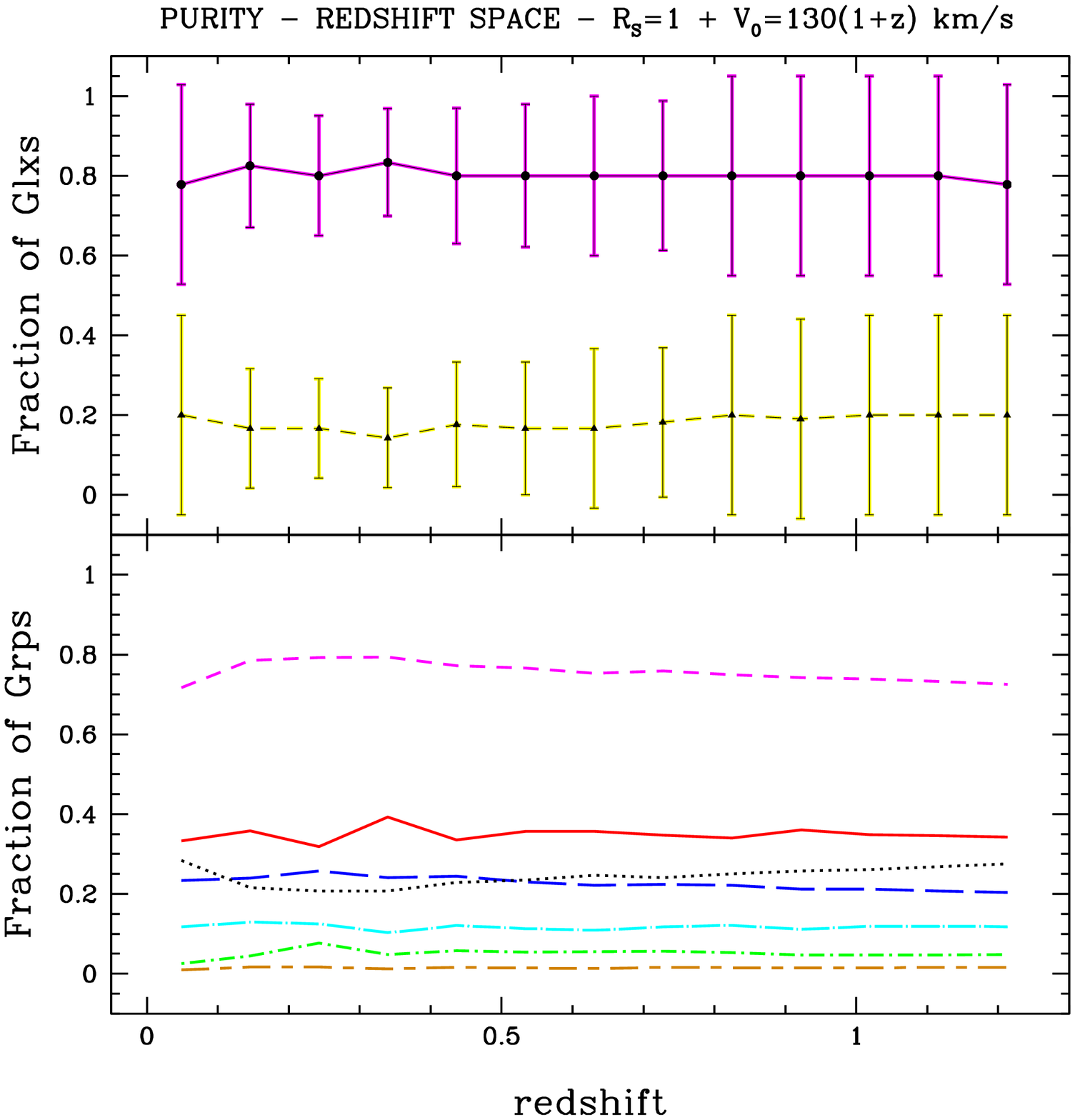}}
{\includegraphics[scale=0.40,clip=]{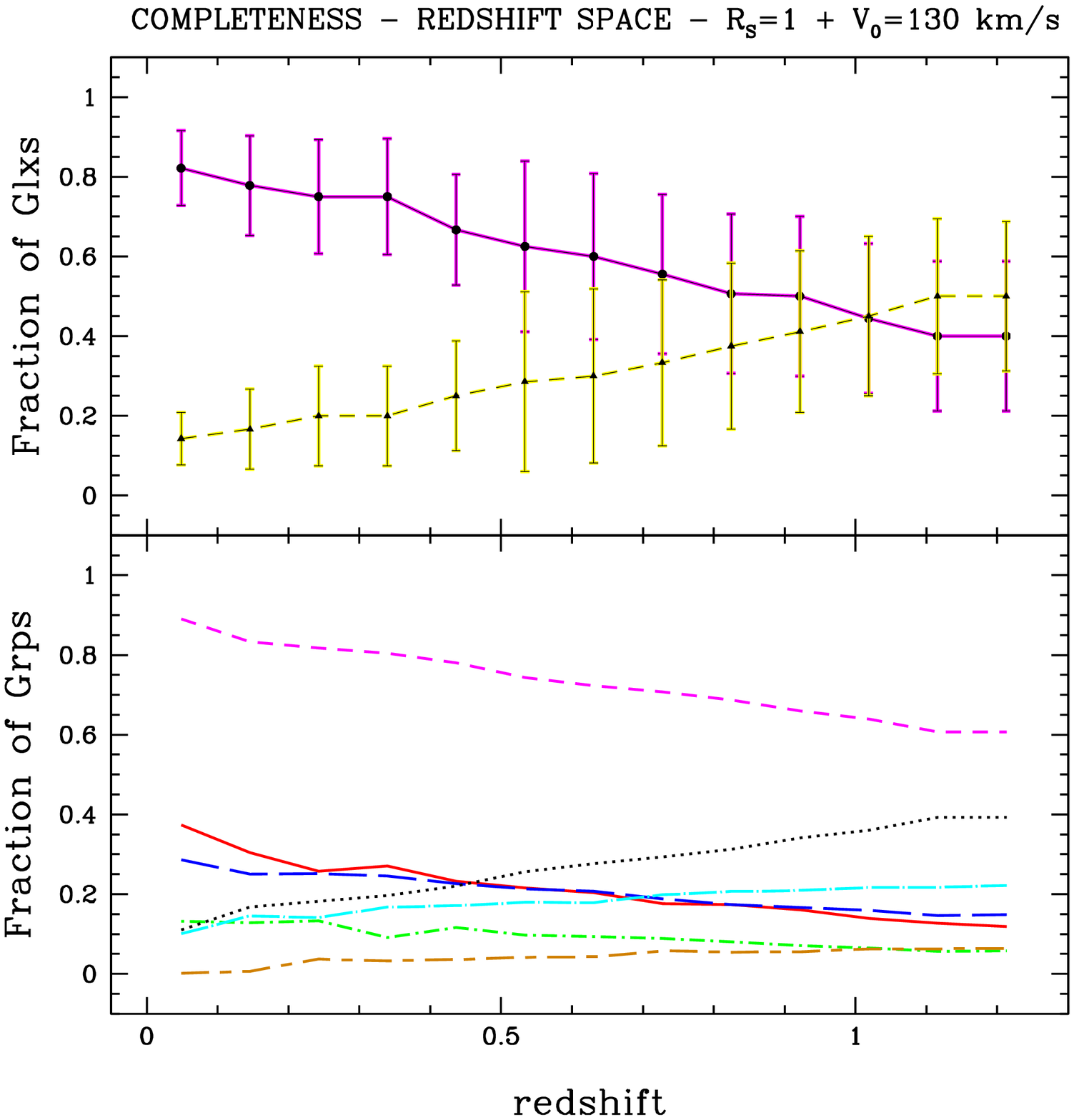}
\includegraphics[scale=0.40,clip=]{c_label.eps}
\includegraphics[scale=0.40,clip=]{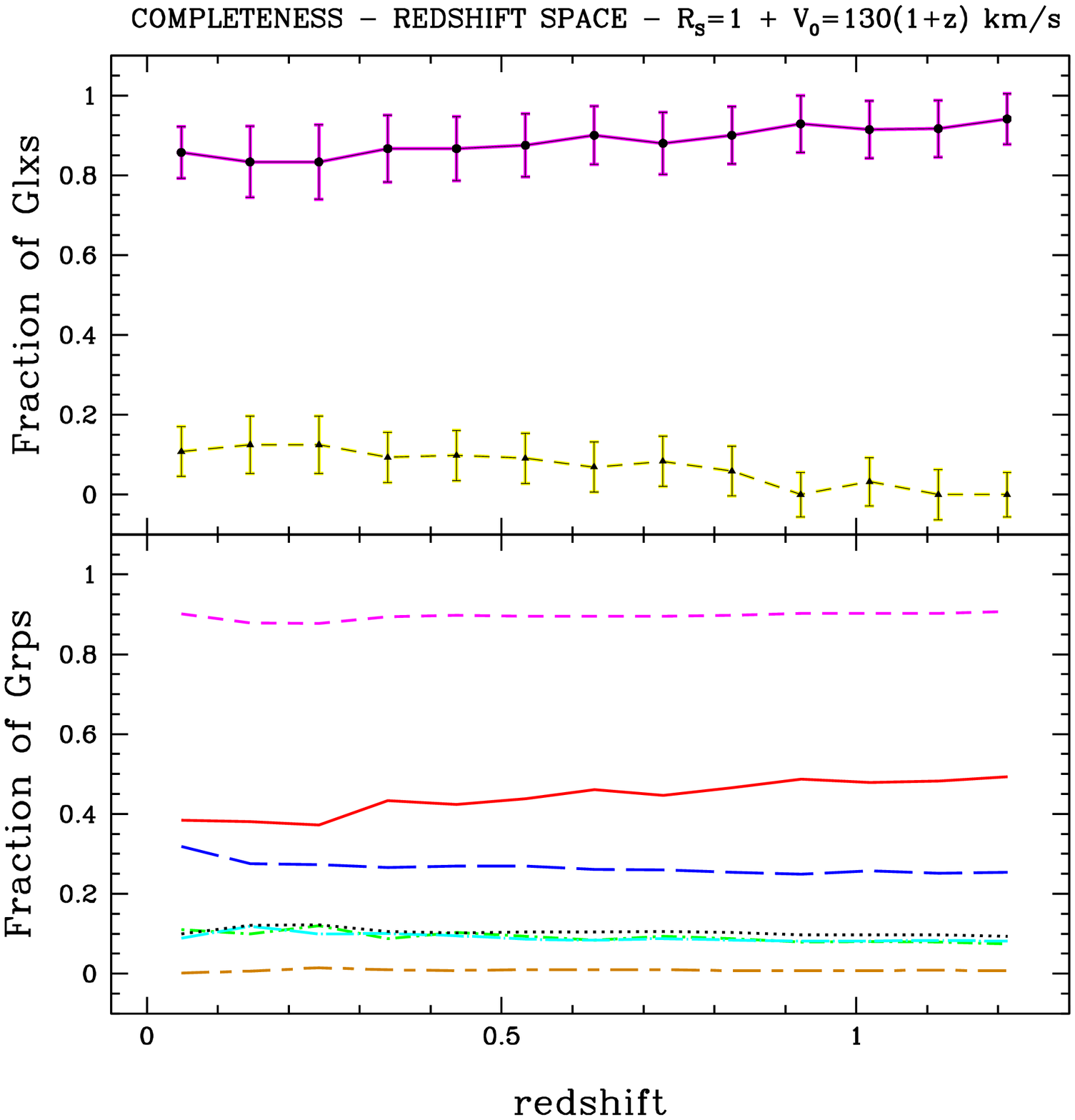}}
\caption{Same as Fig.\ref{purity_flux} but for the samples of groups identified in a volume limited sample 
in redshift space using $V_0=130$ km/s (\emph{left boxes}) or $V_0=130 \, (1+z)$ km/s (\emph{right boxes}) 
\label{purity_red_v1}}
\end{figure*}
\begin{figure*}
\centering
{\includegraphics[scale=0.4,clip=]{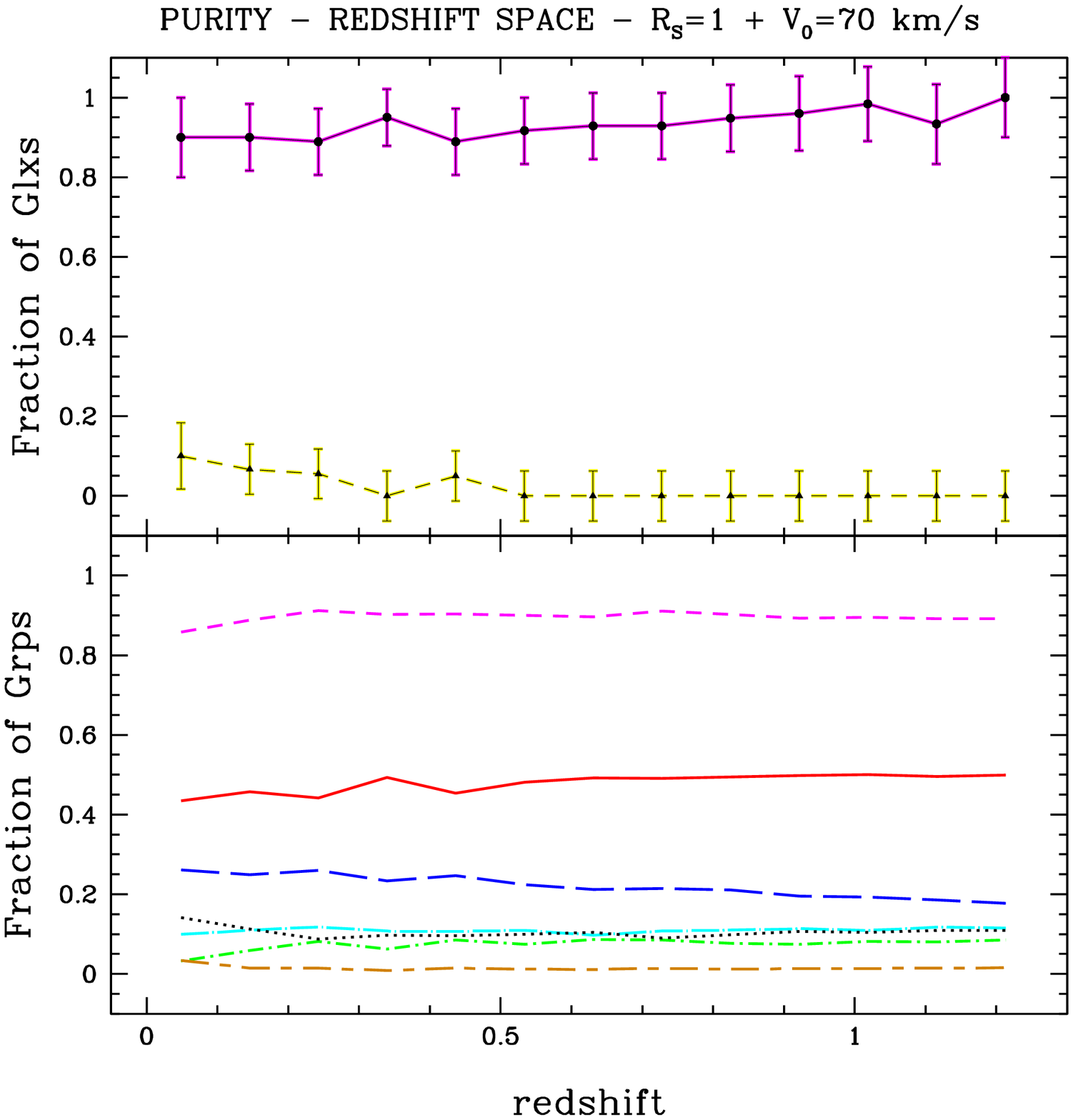}
\includegraphics[scale=0.4,clip=]{p_label.eps}
\includegraphics[scale=0.4,clip=]{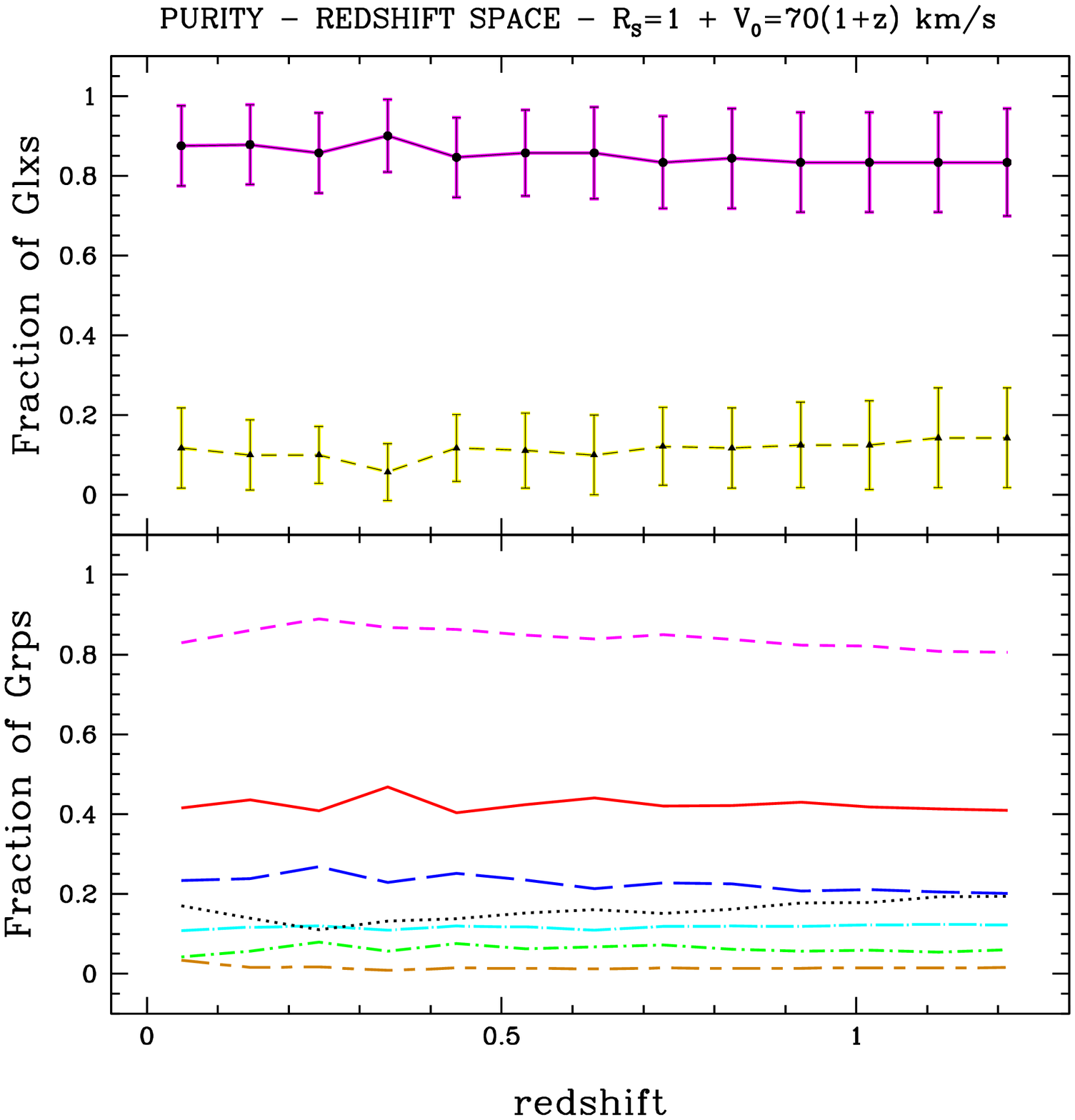}}
{\includegraphics[scale=0.4,clip=]{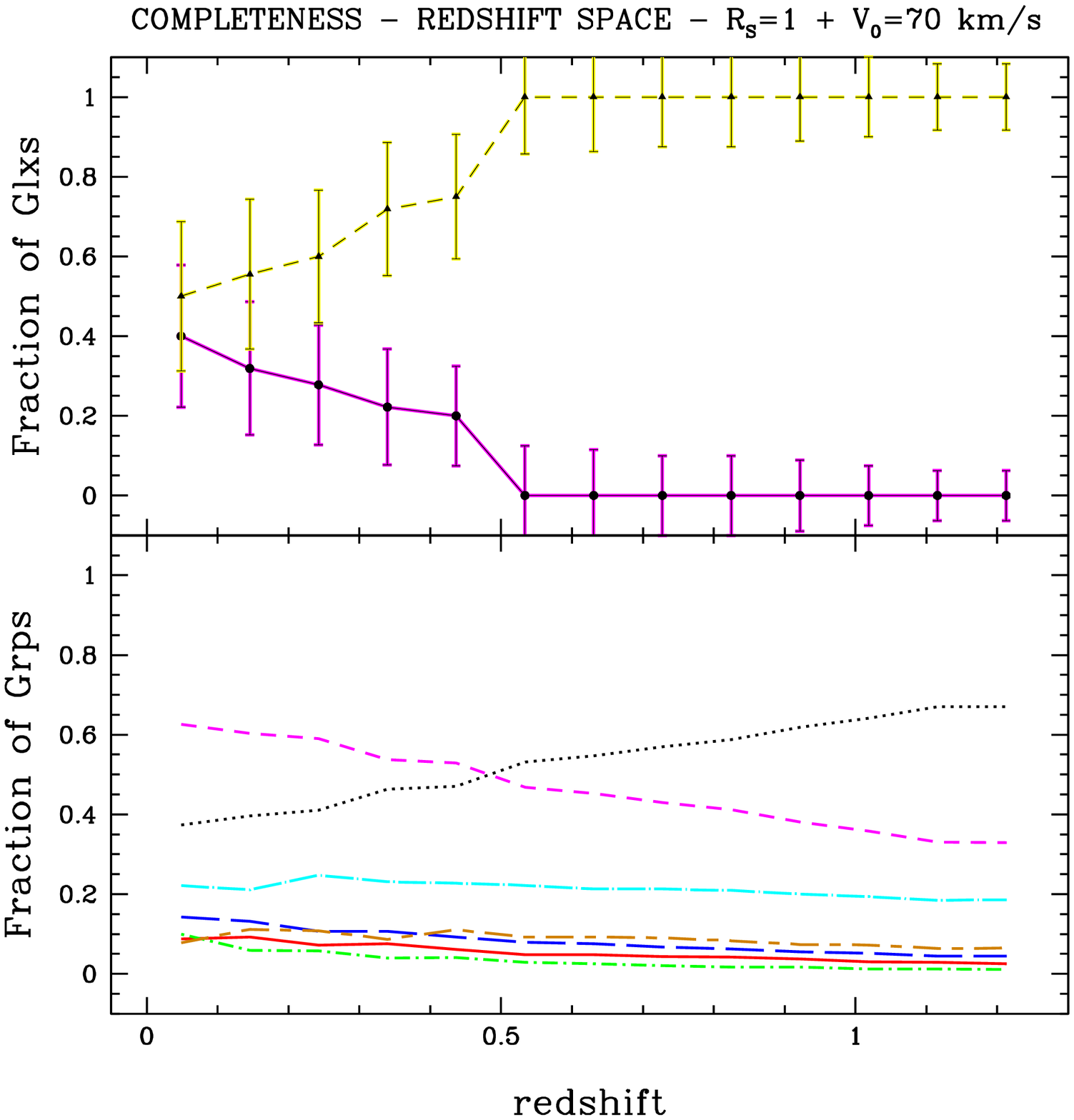}
\includegraphics[scale=0.4,clip=]{c_label.eps}
\includegraphics[scale=0.4,clip=]{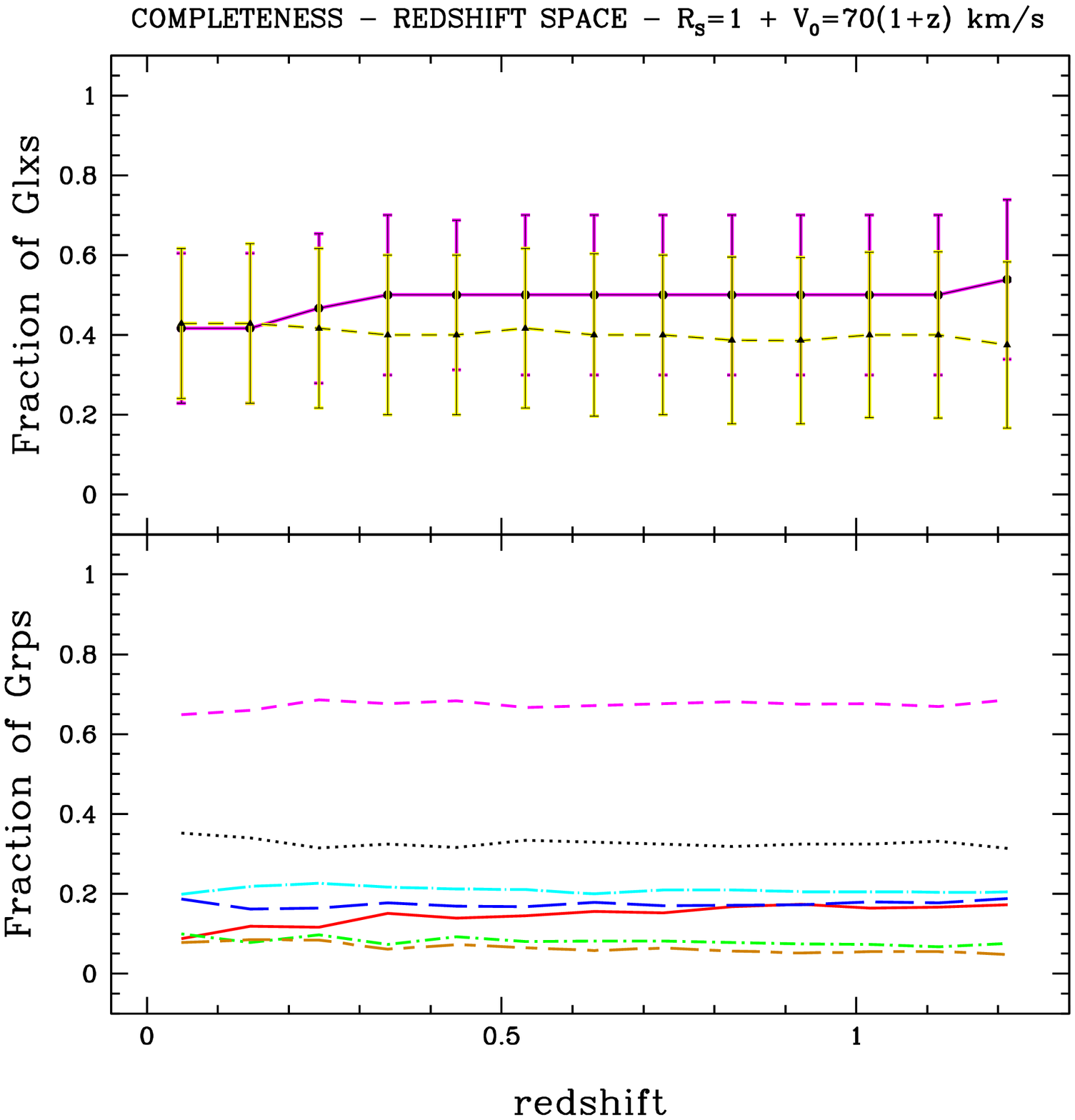}}
\caption{Same as Fig.\ref{purity_flux} but for the samples of groups identified in volume limited sample 
in redshift space using $V_0=70$ km/s (\emph{left boxes}) or $V_0=70 \, (1+z)$ km/s (\emph{right boxes}) 
\label{purity_red_v2}}
\end{figure*}
\begin{figure*}
\centering
{\includegraphics[scale=0.4,clip=]{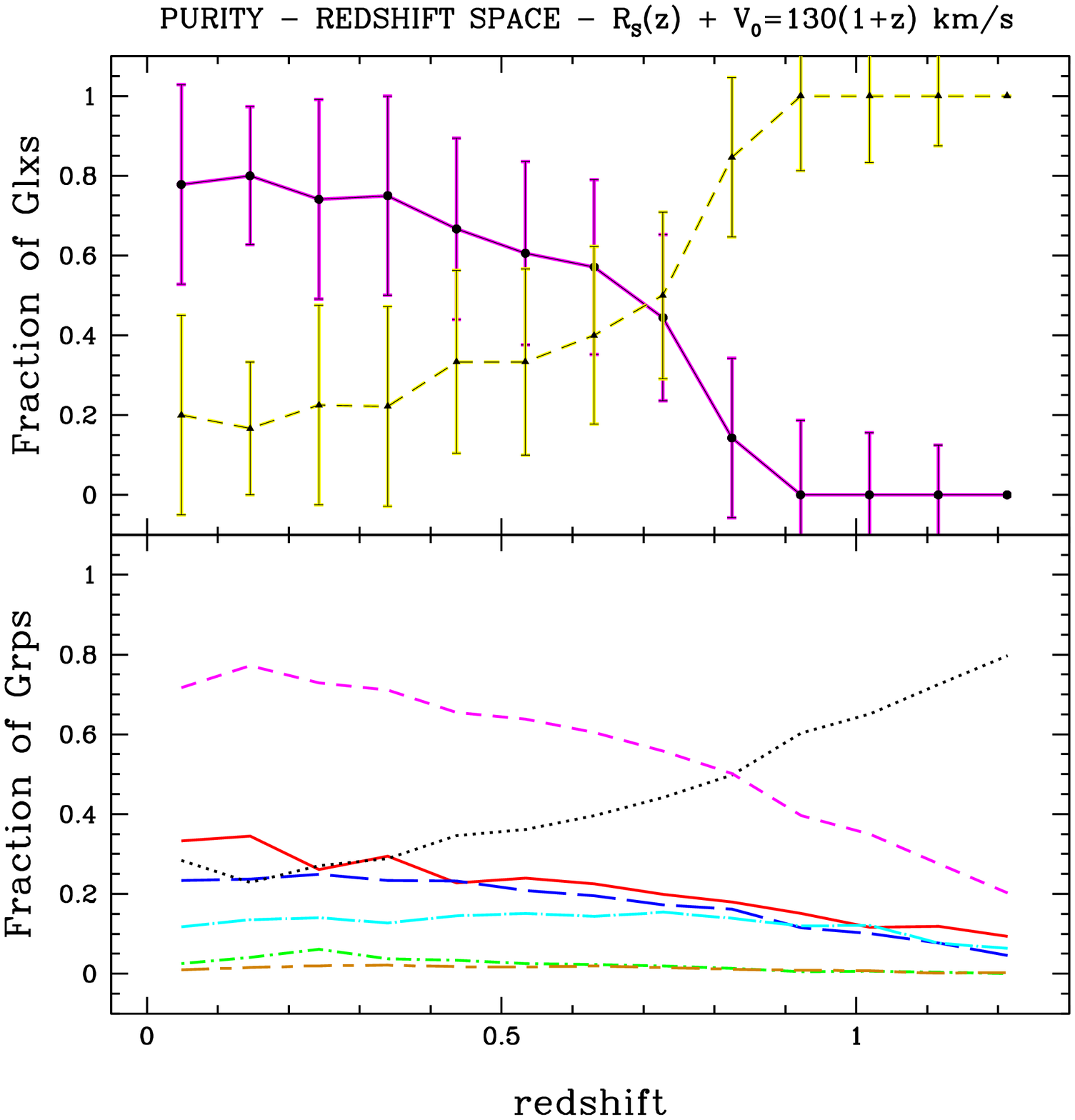}
\includegraphics[scale=0.3,clip=]{p_label.eps}
\includegraphics[scale=0.4,clip=]{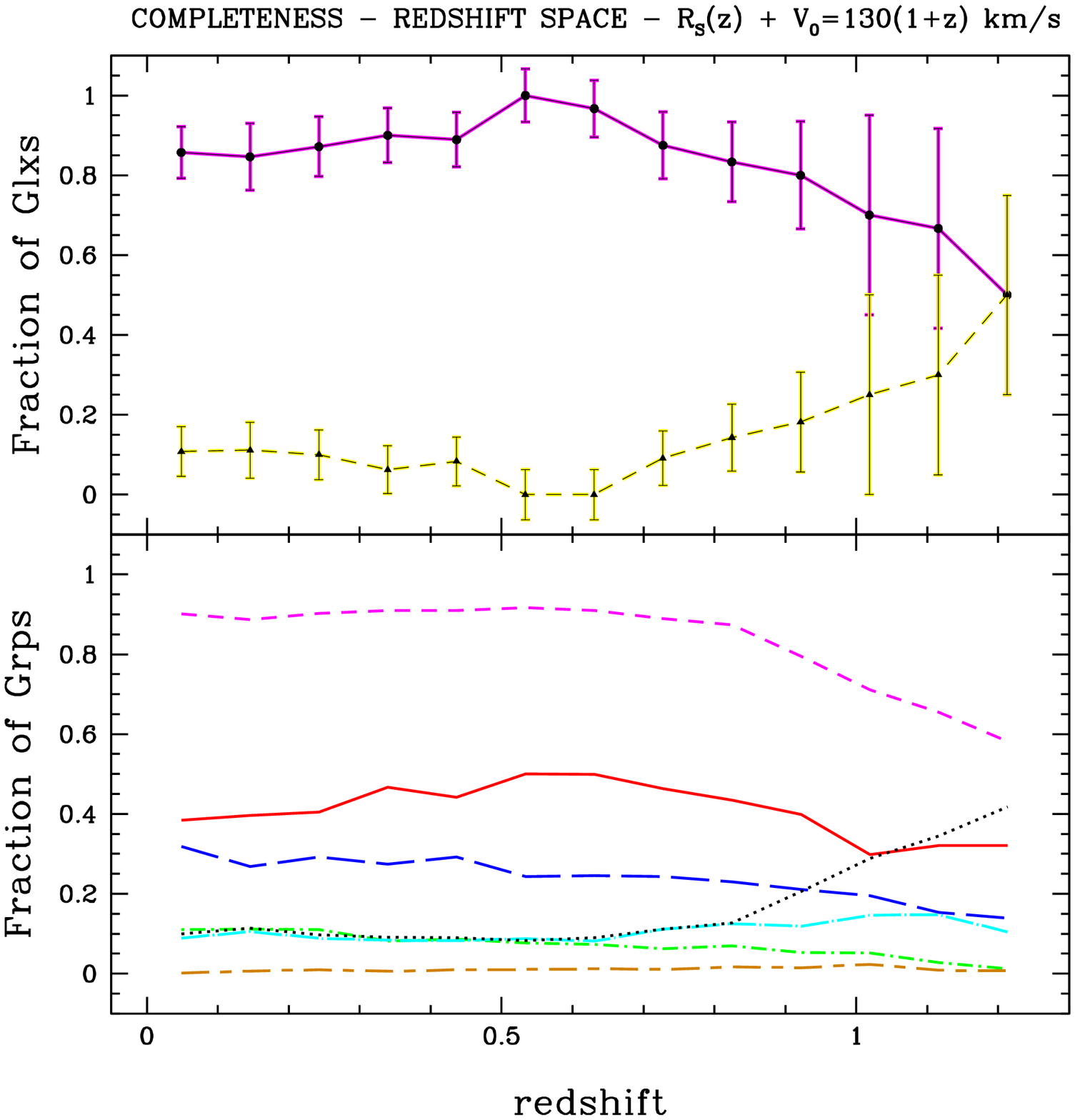}
\includegraphics[scale=0.3,clip=]{c_label.eps}}
\caption{Purity and completeness as a function of redshift for groups identified
in a \emph{flux limited} sample in redshift space. The \emph{left box} shows the purity while 
\emph{right box} shows the completeness when a LF varying with redshift and $V_0=130(1+z)$ km/s 
are used to compute the linking length parameters. 
In the left box (\emph{purity}), the top panels show the fraction of galaxies 
identified that can be associated with galaxies of the group with the highest matching rate
in the corresponding control sample (\emph{solid lines}), 
and the fraction of galaxies that are classified as interlopers (\emph{dashed lines}).
In the right box (\emph{completeness}), the top panels show the fraction of galaxies in the control 
sample that can be associated with galaxies identified in the spectroscopic group 
with the highest matching rate (\emph{solid lines}), 
and the fraction of galaxies from the control sample that are classified as missing (\emph{dashed lines}).
The \emph{bottom panels} show the trends observed for the fraction of groups within 
the six categories of \emph{purity} or \emph{completeness} (see text for 
description). The \emph{magenta short dashed lines} correspond to the complement of $P6$ and $C6$. 
\label{final_sp_sample}}
\end{figure*}

The other observational constraint that needs to be addressed in order to choose the best 
linking length parameters is the redshift space distortion. 
It is necessary to modify the radial linking length $V_l$ when working in redshift space,
since the structures seem elongated along the line of sight due to the infall of galaxies in virialised
galaxy groups. These elongated structures are commonly called Fingers of God.
Therefore, we built a volume limited sample complete down to $i_{SDSS}$ absolute magnitude
$-16.4$, just like the \emph{reference} sample but in this case the positions of galaxies are 
distorted according to eq.~\ref{zdistorted}. The linking length parameters are:
\[ D_l=D_0 \, R_s \ \ {\rm and} \ \ V_l=V_0 \, R_s \] 
The value of $D_0$ is defined above for the \emph{reference} sample, 
the value $R_s$ is taken equal to 1, since there is no flux limit, 
while here we investigate different options for 
the value of $V_0$. Usually, for low redshift samples, $V_0$ is defined as a constant that 
is tuned to produce the more reliable sample of groups in terms of \emph{purity} and/or 
\emph{completeness}.

To determine the best value of $V_0$ in this work, 
we analysed the sample of \emph{reference} groups, and computed the velocity differences in 
redshift space along the line of sight among the group members. 
Our goal is to find the most appropriate value that satisfies the requirement of being the
minimum velocity value needed to link most of the galaxy members of a given group in redshift 
space. Therefore, for each group we looked for the maximum velocity difference 
of the members in the line-of-sight to their closest neighbours. 
These maximum values are shown in the \emph{left upper} panel of Fig.~\ref{delta_v}. 
Dots represent the median values per bin of redshifts while the error bars are their 
semi-interquartile ranges. It can be seen that the maximum velocity difference to the closest 
neighbour is increasing towards higher redshifts. 
In the \emph{left lower} panel, we divided the y-axis by $(1+z)$. 
The medians of these points determine a roughly constant value of $130$ km/s (\emph{solid} line). 
Hence, we are going to test the identification algorithm against using a constant value 
of $130$ km/s and a value that varies with redshift as $130 (1+z)$ km/s. 
Moreover, in order to test the influence of the choice of $V_0$, we also examined a second value. 
Instead of looking for the maximum of the velocity differences to the closest neighbours, 
we also investigated the second maximum of those differences. 
The results are shown in the \emph{right} panels of Fig.~\ref{delta_v}. 
In this case, the values of $V_0$ to be analysed are 
$70$ km/s and $70 (1+z)$ km/s. This second approach, 
with a lower value for $V_0$, is made in order to test whether a 
lower value could improve the resulting group sample in both, \emph{purity} and 
\emph{completeness}.

Therefore, we performed four different identifications. We find $183,717$ groups with more than 4 members
 when using $V_0=130$ km/s, and when $V_0=130(1+z)$ km/s, we find $250,532$.  
With the shorter linking length, we identify $106,920$ and  $197,290$ groups,  
with $V_0=70$ km/s and $V_0=70(1+z)$ km/s respectively (see Table~\ref{grupos}).

As in the previous subsection, we analyse and compare the \emph{purity} and 
\emph{completeness} of these samples in order to choose the best radial linking length parameter. 
The \emph{purity} is defined considering the members of the newly \emph{redshift-space}
identified groups (four samples ${\cal A}$) in comparison with the \emph{reference} sample 
(control sample ${\cal B}$); 
while completeness is defined taking the members in the \emph{reference} sample 
(control sample ${\cal A}$) and looking for their counterparts in the \emph{redshift-space} groups 
(four samples ${\cal B}$). 
The results as a function of redshifts are shown in Figs.~\ref{purity_red_v1} 
and~\ref{purity_red_v2}.

The effect of using either a constant or variable value of $V_0$ can be seen by comparing 
the \emph{left} to the \emph{right} boxes of these figures. 
Firstly, analysing the \emph{purity} in Fig.~\ref{purity_red_v1},
it can be seen that the \emph{purity} of the groups is poorly affected, i.e, 
modulating the linking length by $(1+z)$ or keeping it constant, produces similar results as a function of redshift. 
We observe that roughly 
$\sim 80\%$ of galaxies are associated with the group in the \emph{reference} sample with 
the highest matching rate, while $\sim 20\%$ of galaxies are interlopers. 
From the six-category analysis,for both identifications we find  
$\sim 40\%$ and $\sim 20\%$ of $P1$ and $P2$ groups, respectively. 
When using a constant $V_0$, there are $\sim 20\%$ of misidentified groups ($P6$) 
for any redshift, while this percentage is slightly higher when using a variable $V_0$.

Now, when including the \emph{completeness} analysis for both identifications, 
remarkable differences arise.
For the constant $V_0=130$ km/s, the fraction of galaxies
in the \emph{reference} sample associated with the group in the \emph{redshift-space} sample
with the highest matching rate drastically dropping as a function of redshift, declining to as low as 
$40\%$ at higher redshifts (top panels of left bottom box 
of Fig.~\ref{purity_red_v1}). Also at high redshifts, the $C6$ groups (completely missing) 
reach  $40\%$ and the contribution of $C4$ is $\sim 20\%$ in the whole redshift range.
 On the other hand, when analysing the \emph{completeness}
of the sample identified with variable $V_0$, 
we observe that more than $\sim 80\%$ of galaxies in the \emph{reference}
sample are recovered at all redshifts, with only $15\%$ galaxies missing 
(top panels of right bottom box).
Moreover, the \emph{completeness} is highly improved obtaining 
$\sim 50\%$ of $C1$ groups and more than $\sim 20\%$ of $C2$ groups, and less than $20\%$ of 
missing groups at all redshifts.

From Table~\ref{percentage}, based on the combined percentages of the classes 1 and 2, 
it can be seen that while the percentage of highly pure groups for the identification 
performed with $V_0$ variable is $\sim 7\%$ lower, the percentage of highly complete groups of 
this sample is significantly higher ($+43\%$). Therefore, the best choice for the radial linking length 
is such that it varies with redshift.

By comparing Figs.~\ref{purity_red_v1} and~\ref{purity_red_v2}, it can be seen the effect of 
the amplitude of $V_0$.  Using fixed or variable $V_0$ with $70$ km/s, all the fractions observed 
in the \emph{purity} analysis are slightly higher than those observed and described above 
when using $130$ km/s.
Then, a shorter radial linking length (Fig~\ref{purity_red_v2}) seems 
better in terms of \emph{purity}, i.e, it is able to identify more groups whose members belong 
to some \emph{reference} group ($\sim 7\%$ higher in the total percentage of $P1$ class for both, 
constant or variable $V_0$, see Table~\ref{percentage}). 
However, the results from the \emph{completeness} analysis help choosing the appropriate value. 
For both of the $70$-identifications, the resulting samples are highly incomplete regardless 
the redshift. In the best scenario (considering variable $V_0$), 
the fraction of \emph{reference} members that are included in the \emph{redshift-space} 
groups reaches only $50\%$.
This result implies that shortening the size of the radial linking length 
makes the algorithm to identify fewer of the true groups,
resulting in a completeness for the sample which is quite low. This result is clearer when inspecting
the total percentages of classes $1+2$ in Table~\ref{percentage}. By analysing the identification
with variable $V_0$, it can be seen that the percentage of $C1+C2$ groups drastically drops from  
$73\%$ obtained for $130(1+z)$ to $35\%$ for $70(1+z)$. 
Even more, the resulting group samples obtained when using $70$ km/s are not only incomplete, 
but dominated by groups of category $C6$, the least desired. 

It has also been corroborated that using a value higher than $130$, 
besides not having any physical motivation,
increases the completeness of the sample at the cost of the purity, making it lower than $50\%$.

Therefore, our choice for the radial linking length in redshift space catalogues is 
$V_0=130 (1+z)$ (\emph{right plots} of Fig.~\ref{purity_red_v1}). 
The redshift space distortions make it difficult to recover ``perfectly matched'' groups
($P1$ and $C1$), although they are the most common categories that we identify at all redshifts, 
followed by $P2$ and $C2$. There are $30\%$ of false groups, while the algorithm is not able to 
recover only $10\%$ of the true groups.  
All in all, the resulting sample has more than $50\%$ of highly pure groups
while we are able to identify $73\%$ of the highly complete groups.

\subsection{Spectroscopic sample: flux limited sample in spectroscopic-redshift space} 
\label{sp-sample}
After having chosen the best linking length parameters, we identify groups in the mock galaxy 
catalogue described in Sect.~\ref{mock}. 
The identification is therefore performed with the following linking lengths:
\[ D_l=D_0 \, R_s \ \ {\rm and} \ \ V_l=130 (1+z) \, R_s \]
with $D_0$ and $R_s$ defined in eqs.\ref{d0} and~\ref{rs}, respectively, and 
using a variable luminosity function.

The algorithm produces a sample of $23,183$ \emph{mock} groups with 4 or more members 
(see Table~\ref{grupos}). The \emph{purity} and \emph{completeness} as a function of redshifts for
this sample are shown in Fig~\ref{final_sp_sample}. Both statistics are computed using the
\emph{restricted-reference} groups as the control sample.
It can be seen the combined effect of both observational constraints, the flux limit and the redshift 
space distortions. Regarding the \emph{purity}, the fraction of members in the spectroscopic groups 
that also belong to the \emph{restricted-reference} group with the highest matching rate 
(\emph{top panels in the left box}) drastically decreases towards higher redshifts, 
ranging from $\sim 80\%$ to $\sim 0\%$. 
When analysing the six categories of groups defined above, 
an increase in false identification ($P6$ groups) can be seen towards higher redshifts, with the 
sample at redshifts higher than $z=0.8$ being dominated by these false groups. The ``perfectly matched'' groups 
($P1$) and ``quasi-perfectly matched'' groups ($P2$) are the more frequent among the other categories. 
Groups associated with a single real group plus more than $30\%$ of interlopers ($P4$) represent 
$\sim 10\%$ at all redshifts. 

From the \emph{completeness} analysis (\emph{right box}), 
the fraction of members in the \emph{restricted-reference} sample that we have been able to 
identify in the spectroscopic group with highest matching rate (\emph{top panels}) 
decreases with redshift, i.e, it is more likely to lose some of the true members at high redshift.

The ``perfectly recovered'' groups ($C1$) are dominant at all redshifts, followed by those groups where 
only a few members are missing ($C2$). The fraction of completely missing groups is almost constant at
 $\sim 10\%$ up to $z=0.8$, and then increases towards higher redshifts.  

\begin{figure}
\centering
{\includegraphics[scale=0.4,clip=]{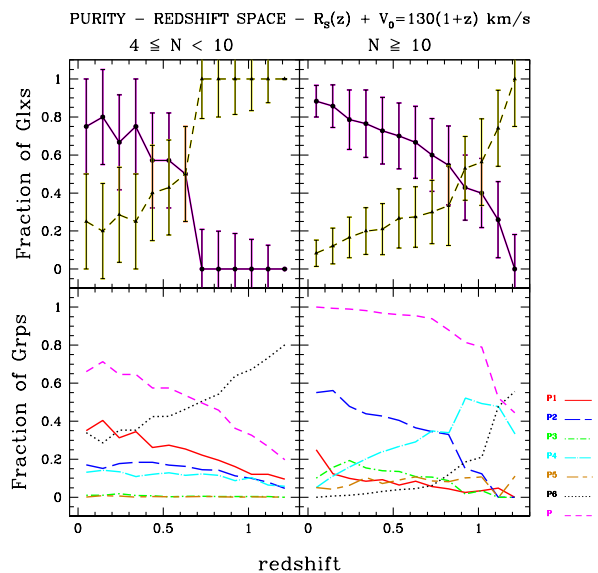}
\includegraphics[scale=0.3,clip=]{p_label.eps}}
\caption{Purity as a function of redshift for groups identified in a spectroscopic catalogue using a LF varying with redshift and $V_0=130(1+z)$ km/s to compute the linking length parameters. 
\emph{Left panels} correspond to low membership groups
while \emph{right panels} are the high membership ones.
The \emph{top panels} show the fraction of galaxies 
identified that can be associated with galaxies of the group with the highest matching rate
in the corresponding control sample (\emph{solid lines}), 
and the fraction of galaxies that are classified as interlopers (\emph{dashed lines}).
The \emph{bottom panels} show the trends observed for the fraction of groups within 
the six categories of \emph{purity} (see text for description). 
The key for colours and line types is the same as in the previous figure. 
\label{spectro_num}}
\end{figure}

To deepen our study, we analysed the purity of the \emph{spectroscopic} groups splitting the sample 
into low ($<10$) and high ($\ge 10$) membership groups. The results are shown in Fig.\ref{spectro_num}. 
It can be seen that the low membership groups are more prone to include false identifications ($P6$), 
while this category is almost non-existent at low redshifts among the high membership groups,
and it increases towards higher redshifts. 
The ``perfectly-matched'' groups are not frequent in the high membership groups, 
however this sample is dominated by the ``quasi-perfectly-matched'' groups until $z=0.7$,
 and groups with more than $30\%$ of interlopers ($P4$). The $P3$ groups (merging) are $\sim 20\%$ at al redshifts. 
These results indicate that the low membership group sample is highly contaminated, 
and it is strongly recommended not to use it for statistical purposes. 

Analysing the total percentages within each of the purity and completeness classes (Table~\ref{percentage}),
we find that the spectroscopic group catalogue has $41\%$ of groups of high-quality purity ($P1+P2$), 
while the $69\%$ of the \emph{restricted-reference} sample is well recovered ($C1+C2$). The false groups 
($P6$) sum up to $\sim 43\%$, mainly due to low membership false groups, while we completely 
loose $\sim12\%$ of the true groups ($C6$). 
A closer inspection to the \emph{lower panels} of Fig.~\ref{spectro_num} reveals that at low redshifts 
the percentage of false groups is lower than $40\%$ for low membership groups,
while it is negligible for high membership groups, which means that our choices of the linking lengths
produce similar results to those that were found in low redshift catalogues by \cite{merchan02}.

\subsection{Photometric sample: flux limited sample in photometric-redshift space}. 

In this section we perform a similar analysis than in the previous section but focusing in
observational catalogues with distances calculated using only photometric information, i.e.,
by means of photometric redshifts.

\begin{figure}
\begin{center}
\includegraphics[width=90mm]{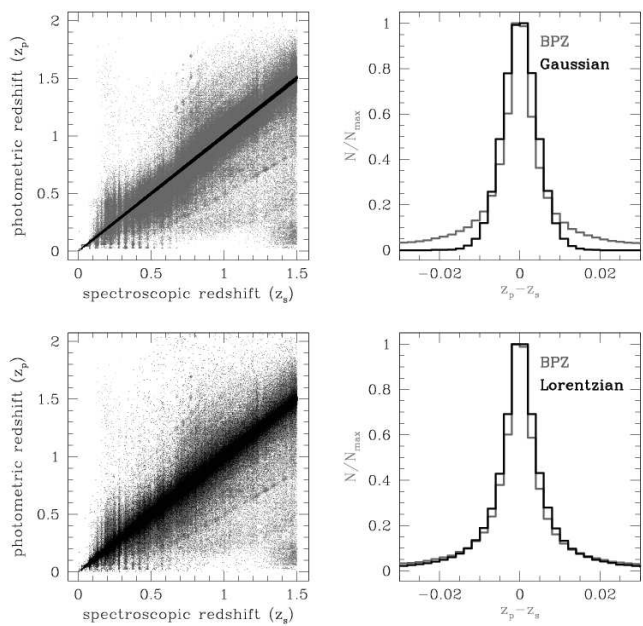}
\caption{Left panels: scatter plots between the spectroscopic redshift ($z_s$) and the photometric
redshift ($z_p$). 
Right panels: close-up of the distribution of the redshift differences $z_p-z_s$ for $|z_p - z_s| < 0.03$. 
The grey colour is used when the photometric redshifts are computed using the
BPZ code, while the black colour is used when the photometric redshifts are assigned randomly. In the upper
panels the random photometric redshifts are assigned using a Gaussian probability 
distribution with the spectroscopic redshift as the centre and 0.0025 as the Gaussian width. 
In the lower panels the random photometric redshifts are 
assigned using a Lorentzian probability distribution, with a width of 0.00244 (see text for full description). 
}
\label{photozscat}
\end{center}
\end{figure}

\subsubsection{The probability Friend-of-Friends: PFOF}
To take into account the uncertainties of using photometric redshifts, we modify the identification
algorithm in the line of sight direction using the method developed by \cite{liu08}.

Instead of just computing the module difference among the velocities of a galaxy 
pair ($\left| V_2-V_1 \right|$) and restricting it to be smaller than $V_l$, 
the definition of a galaxy pair has to take into account the probabilistic nature of 
the photometric redshifts, and therefore the algorithm has to compute 
the probability of the distance between two galaxies 
to be less than the linking length, and then restrict such probability with an artificial threshold. 
Therefore, following \cite{liu08}, the probability 
of two galaxies being closer than $V_L$ is:
\begin{equation}
P(\left| V_2-V_1 \right| \leq V_l) \equiv \int_{0}^{\infty} dz \ F_1(z) \int_{z-V_l}^{z+V_l} 
dz' \ F_2(z')
\label{prob_eq}
\end{equation}
where $F_1$ and $F_2$ are the probability distribution functions for the two galaxies in the line of
sight direction. Therefore, the line of sight criterion to determine that two galaxies are physically 
associated is
$$P(\left| V_2-V_1 \right| \leq V_l) > P_{th}$$
where $P_{th}$ is an appropriate probability threshold. This threshold will be determined 
in the sections below in order to obtain a sample of groups with the suitable balance between
purity and completeness. 

\subsubsection{Testing the PFOF algorithm}
In order to apply this modification to our algorithm, we have to adopt a probability distribution 
for the photometric redshifts. 

The most common model used in the literature when working with 
photometric redshifts is a Gaussian probability distribution \citep{liu08,ascaso12}. Therefore, 
we follow that approach and model the probability distribution
function associated with each galaxy by a Gaussian function, i.e.:
$$G_i(z)=\frac{1}{\sigma_i \sqrt{2\pi}} \exp{\left( \frac{-(z-z_i)^2}{2{\sigma_i}^2} \right)}$$
where $z_i$ is the photometric redshift and $\sigma_i$ the photometric redshift error
of galaxy $i$.

But also, we adopt a different probability distribution, a Lorentzian function, and test the 
behaviour of the method against different distributions. A Lorentzian function is given by:
$$L_i(z)= \frac{1}{\pi \sigma_i} \frac{1}{1+(\frac{z-z_i}{\sigma_i})^2} $$

\begin{table}
\begin{center}
\tabcolsep 5pt
\caption{Groups identified in different galaxy samples in photometric-redshift space \label{grupos_photoz}}
\begin{tabular}{lcrrr}
\hline
Probability  &  $P_{th}$  & Gals in & Groups with   & Groups with \\ 
function   &            & groups  & $4\le N < 10$ & $N \ge 10$  \\
\hline
Random photo-z \\
\emph{Gaussian} &  99 & 40104 & 4958 & 704 \\
&  95 & 99035 & 9974 & 1849 \\
&  90 & 137463 &12727 & 2750  \\
 &  80 & 186467 &16435 & 3931 \\
 &  70 & 220373 & 19213& 4710  \\
 &  60 & 246750 & 21457& 5246  \\
 &  30 & 306589 & 25631& 6601  \\
\emph{Lorentzian} &  99 & 22304 & 3098 & 342 \\
 &  95 & 61124 & 6863 & 1107 \\
 &  90 & 89776 & 9426 & 1688 \\
 &  80 & 131329& 13071& 2595 \\
&  70 & 164875& 15882& 3365 \\
&  60 & 195820& 18417& 4036 \\
&  30 & 300396& 25407& 6346 \\
 & \\
BPZ photo-z \\
\emph{Gaussian} &  99 & 7200 & 1267 & 70 \\
&  95 & 34875 & 4658 & 562 \\
 &  90 & 56404 & 6960 & 987 \\
&  80 & 86326 &10008 & 1567 \\
&  70 & 111020&12517 & 2020 \\
&  60 & 132277&14476 & 2475 \\
&  30 & 194463&19965 & 3755 \\
\emph{Lorentzian} &  99 & 7126 & 1256 & 68 \\
&  95 & 34828 & 4653 & 561 \\
 &  90 & 57020 & 7031 & 989 \\
 &  80 & 90147 & 10421 & 1609  \\
 &  70 & 119251 & 13335 & 2177 \\
 &  60 & 146656 & 15856 & 2780  \\
 &  30 & 251191 & 23899 & 4984  \\

 \hline
\end{tabular}
\end{center}
\end{table}

Firstly, we test the PFOF algorithm in the case where the galaxy redshifts have small uncertainties,
as it is true in the case of spectroscopic redshifts. We adopt $\sigma_i=30$ km/s (the typical 
error in SDSS), and apply the PFOF to the mock galaxy catalogue described in Sect.~\ref{mock} 
using a Gaussian probability distribution in Eq.~\ref{prob_eq}. 
We identify $23,239$ groups having 4 or more members using a probability threshold of $99\%$.
Choosing as control sample the groups identified in Sect.~\ref{sp-sample}, 
the analyses of completeness and purity
revealed that the new identification is $99\%$ pure and $99\%$ complete, considering just 
the combined fractions $P1+P2$ and $C1+C2$, defined in the previous sections.
This means that, in the limit of small uncertainties, 
the PFOF algorithm behaves as the original FOF algorithm.

As a second test, the value of $\sigma_i$ is adopted in order to mimic the difference between the 
BPZ photometric redshifts and the spectroscopic redshifts shown in the \emph{upper right panel} 
of Fig.~\ref{photozscat} (\emph{grey histogram}).  
Choosing a Gaussian function to fit the differences, 
we adopt as the best fitting\footnote{we used the Levenberg-Marquardt method to fit no-linear functions} 
redshift error $\sigma_i=0.0025 (1+z_s)$ for all galaxies. 
We also adopt a Lorentzian probability distribution to fit the differences. 
The best fitting redshift error for the Lorentzian function is $\sigma_i=0.00244 (1+z_s)$.

Then, we modify the redshifts of the galaxies in the mock catalogue by randomly shifting 
the spectroscopic redshifts according to the previously fitted probability distributions: 
we generate a sample with the gaussian distribution and a sample with the lorentzian distribution.
The distribution of differences for the resulting random samples are shown in Fig.~\ref{photozscat}.
The sample generated with the Gaussian distribution is shown as the \emph{black histogram} in the \emph{upper right panel}. 
It can be seen that this distribution reproduces the mean of that obtained from a more realistic 
determination of photometric redshifts (BPZ). However, it is not possible to reproduce the tails of the 
realistic distribution when using a simple Gaussian function. 
The resulting redshift differences for the random lorentzian sample 
are shown as \emph{black histogram} in the \emph{lower right panel} of Fig.~\ref{photozscat}. In this case, 
the mean and the tails of the original distribution are well recovered.

\begin{figure}
\begin{center}
\includegraphics[width=\hsize]{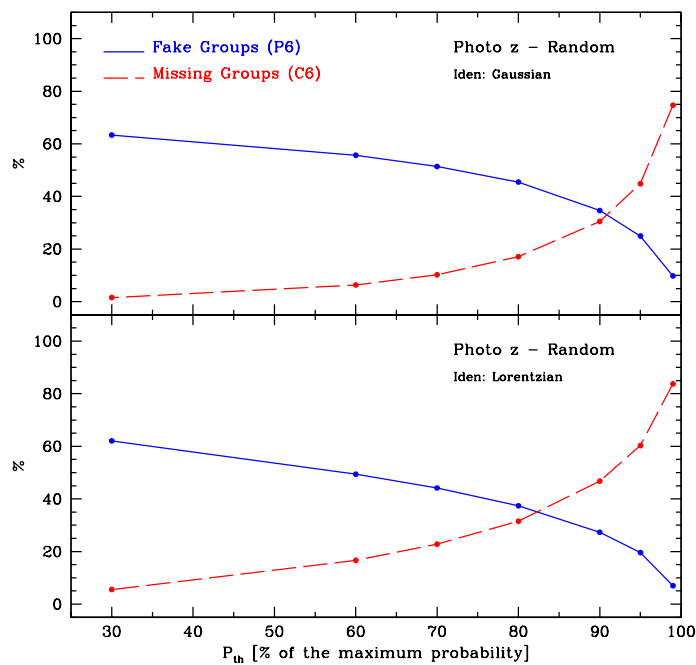}
\caption{Percentages of false groups (\emph{blue solid line}) and missing groups (\emph{red dashed line}) 
as a function of the probability threshold used in the group identification algorithm. 
\emph{Top panel} (\emph{bottom panel}) corresponds to the identifications performed using a Gaussian 
(Lorentzian) function in the PFOF algorithm and in the assignment of random photometric redshifts. 
}
\label{pctotran}
\end{center}
\end{figure}
\begin{figure}
\begin{center}
\includegraphics[width=\hsize]{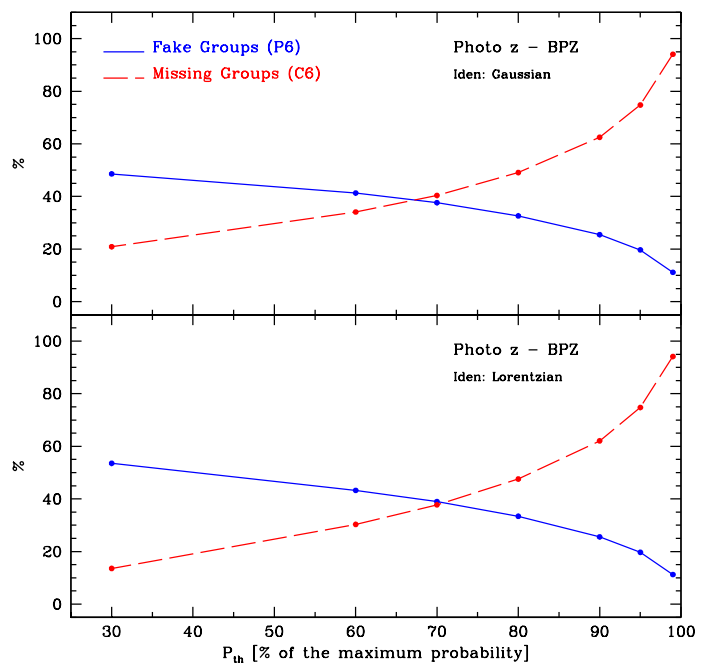}
\caption{Percentages of false groups (\emph{blue solid line}) and missing groups (\emph{red dashed line}) 
as a function of the probability threshold used in the group identification algorithm. 
\emph{Top panel} (\emph{bottom panel}) corresponds to the identifications performed using a Gaussian 
(Lorentzian) function in the PFOF algorithm. The photometric redshifts were assigned using the BPZ code. 
}
\label{pctotbpz}
\end{center}
\end{figure}

We test the PFOF algorithm in both samples, 
one with photometric redshifts generated from a Gaussian function, 
and the other where the photometric redshifts come from a Lorentzian function.  
The application of the PFOF is straightforward, just using for each galaxy
the input distribution from which their redshifts have been generated to compute the probability
of Eq.~\ref{prob_eq}. 

We tested different probability thresholds to perform the identification on the different samples. 
These thresholds are defined as being a percentage (99, 95, 90, 80, 70, 60 and 30\%) of the 
maximum probability obtained from Eq.~\ref{prob_eq}. 
The effect of choosing different thresholds will be described
in the analyses of purity and completeness of the resulting group samples.

The number of groups identified in each sample is shown in Table~\ref{grupos_photoz}.

We analyse the purity and completeness of these samples of groups taking as 
control sample the \emph{restricted-reference} group sample, 
defined in Sect.~\ref{reference-samples}. 
In Fig.~\ref{pctotran} we show the percentage of groups identified with PFOF and 
classified as having purity $P6$ (\emph{blue solid lines}), 
and the percentage of groups of the \emph{restricted-reference} sample that have been lost
 by the PFOF algorithm ($C6$, \emph{red dashed lines}), 
both as a function of the probability threshold. 
We chose to show here only these categories since they show how bad the identification was.
In this figure, the top panel corresponds to the identifications performed on samples of galaxies 
with photometric redshifts assigned randomly according to a Gaussian distribution, while the bottom
panel shows the results for the samples where the photometric redshifts come from a Lorentzian 
distribution. It can be seen that the percentage of false identifications decreases towards higher 
probability thresholds, while the opposite happens with the percentage of the missing groups. 
An appropriate choice of the probability threshold would be the value where both trends overlap, 
i.e., $P_{th}=91\%$ for the Gaussian distributions, and $P_{th}=82\%$ for the Lorentzian distribution. 
Having chosen the probability threshold, in both samples, the false groups will sum up to $~35\%$ 
as well as the missing groups. 

\subsubsection{Application of PFOF to mock galaxies with BPZ photometric redshift}
We now test the PFOF algorithm when applied to mock galaxies whose photometric redshifts have been 
computed in a realistic way (see Sect.~\ref{bpz_method}). 
We identify 2 samples of groups: 
(i) the algorithm works with a probability Gaussian function with $\sigma_i=0.0025 (1+z_p)$, and 
(ii) the algorithm works with a probability Lorentzian function with $\sigma_i=0.00244 (1+z_p)$

The numbers of groups identified for the different probability thresholds are shown in 
Table~\ref{grupos_photoz}.
In order to determine the purity and completeness of these samples, we take as control sample the
\emph{restricted-reference} sample of groups. In Fig.~\ref{pctotbpz}, the percentages of 
false groups ($P6$) groups and the missing groups ($C6$) are shown as a 
function of the probability thresholds. 
The global behaviour of the trends are similar to what we found
when the photometric redshifts were assigned randomly. It can be seen that there is little 
difference in the identifications when using a Gaussian function to describe the 
distribution of the photometric redshifts or a Lorentzian function, 
although the Lorentzian distribution is a better description for the data 
in a wider range (Fig~\ref{photozscat}).
The appropriate probability threshold are $P_{th}=67$ when using Gaussian functions in the algorithm, 
and $P_{th}=70$ when using Lorentzian functions. The percentage of false and missing groups 
are $\sim 40\%$. 

The total percentages of purity and completeness within each category for the different 
probability threshold when using a Lorentzian function in the PFOF algorithm are 
quoted in Table~\ref{total_PC_photoz}. 
$P_{th}=70\%$ is the best compromise to obtain higher percentages of purity and completeness
(or lower fractions of false and missing groups). 
\begin{table}
\begin{center}
\tabcolsep 5pt
\caption{Total percentages of purity and completeness of groups identified in mock galaxy catalogue 
with realistic photometric redshifts \label{total_PC_photoz}}
\begin{tabular}{ccccccccc}
\hline
Class & & \multicolumn{7}{c}{BPZ - Lorentzian - $P_{th}$ }  \\ 
      & & 30 & 60  &  70 &  80 & 90 & 95 & 99\\
\hline
P1   & & \ 2 & \ 5& \ 6 & \ 8 & 12 & 16 & 30  \\
P2   & & \ 7 & 11 & 13 & 16 & 19 & 22 & 23 \\
P3   & & \ 2 & \ 5 & \ 6 & \ 7 & \ 9 & 10 & 11 \\
P4   & & 27 & 28 & 29 & 29 & 28 & 27 & 21 \\
P5   & & \ 8 & \ 7 &  \ 7 & \ 6 & \ 6 & \ 6 &  \ 4 \\
P6   & & 54 & 44   & 39 & 34 & 26 & 19 & 11 \\
\hline
C1   & & 12 &  \ 5 & \ 4 & \ 2 & \ 1 & \ 0 & \ 0 \\
C2   & & 22 & 15 & 11 & \ 8 & \ 3 & \ 1 & \ 0 \\
C3   & & 23 & \ 9 & \ 6 & \ 4 & \ 2 & \ 1 & \ 0 \\
C4   & & 22 & 31 & 32 & 30 & 25 & 17 & \ 4 \\
C5   & & \ 7 & \ 9 & \ 9 & \ 9 & \ 7 & \ 6 & \ 2\\
C6   & & 14 & 31 & 38 & 47 & 62 & 75 & 94  \\
\hline
\end{tabular}
\end{center}
\end{table}

\begin{figure}
\begin{center}
\includegraphics[width=\hsize]{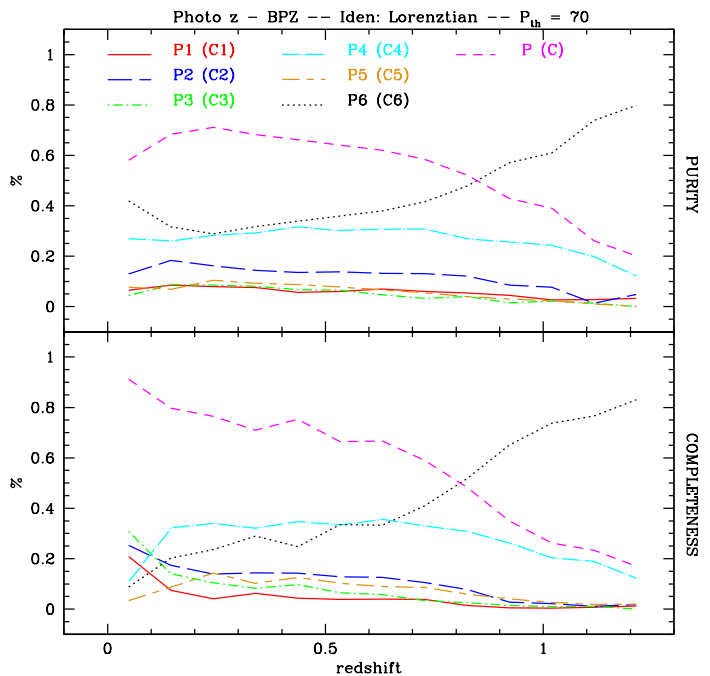}
\caption{\emph{Top panel}: Percentage of photometric groups split
into the six categories of purity as a function of redshift. 
\emph{Bottom panel}: Percentage of \emph{restricted-reference} split into the 
six categories of completeness as a function of redshifts
}
\label{pctotz}
\end{center}
\end{figure}

We also investigate the variation of the fraction of groups within each of the six categories of 
purity and completeness as a function of redshifts. We choose as our main sample that obtained when 
using a Lorentzian function in the PFOF algorithm and a probability threshold of $P_{th}=70$. 
The resulting trends are shown in Fig.~\ref{pctotz}.

Regarding the purity (\emph{top panel}), it can be seen that the resulting sample of 
groups is dominated by false groups ($P6$) at all redshifts, 
followed by groups having less than 70\% of galaxies 
that belong to one true group ($P4$). 
Perfect or quasi perfect matched groups are less than $20\%$ in the whole 
redshift range, which is expectable given the probabilistic nature of the identification. 
In this figure the \emph{magenta short dashed line} represent the sum of all the categories 
but the $P6$, resembling groups that contain at least part of the true groups. 
At redshifts lower than $0.85$ the contribution of all these categories together is higher 
than the contribution of the false groups, while this behaviour reverses at higher redshifts. 

The analysis of the completeness is shown in the \emph{bottom panel} of Fig.~\ref{pctotz}.
Most of the true groups are missing at redshifts higher than $0.8$, 
which is shown with the \emph{black dotted line} ($C6$). At lower redshifts, 
groups with less than $70\%$ of their members identified in the photometric sample are dominant. 
The contribution of all true groups whose members have been included total or partially 
in any photometric group (the sum of all the categories but C6) is higher than $60\%$ at 
redshifts lower than $0.8$.  

\begin{figure}
\begin{center}
\includegraphics[width=\hsize]{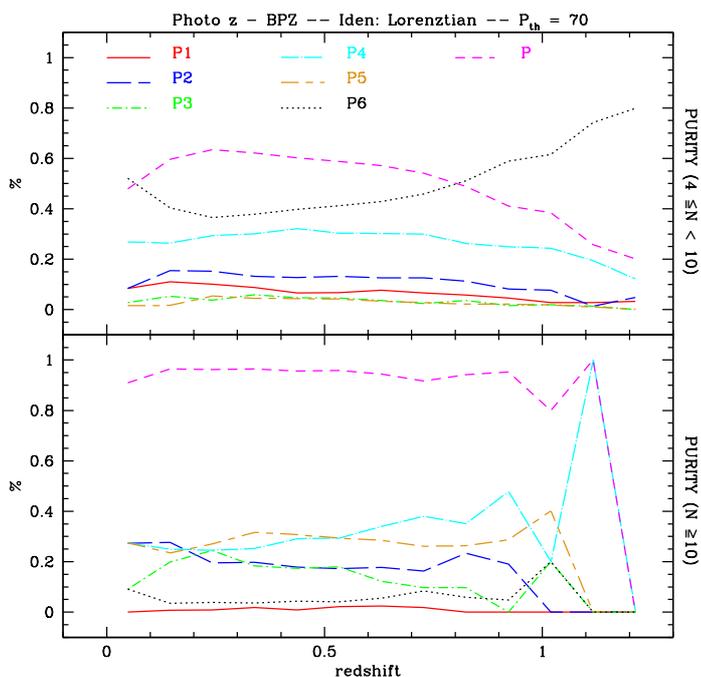}
\caption{Purity of groups identified in a realistic photometric mock galaxy sample.
\emph{Top panel}: Percentage of low membership photometric groups split
into the six categories of purity as a function of redshift. 
\emph{Bottom panel}: Percentage of high membership photometric groups split
into the six categories of purity as a function of redshift. 
}
\label{pctotz_num}
\end{center}
\end{figure}

We also split the sample of photometric groups into low and high membership groups 
(for groups with $ 4 \le N < 10 $ and $N \ge 10$, respectively). 
The six category analysis of purity for low and high membership groups is shown in 
Fig.~\ref{pctotz_num}.
The \emph{top panel} of this figure shows that the low membership groups 
are responsible for the high contamination by 
false groups ($P6$) in the sample in the whole redshift range. 
High membership false groups are less than $10\%$ at all redshifts, indicating that groups 
that contain at least part of the true groups sum up to roughly $90\%$. 
Therefore, we suggest not using the low membership sample identified with this algorithm to perform 
statistical studies.


\section{Summary}
In this work we have performed a detailed analysis in order to assess the reliability of 
a FoF algorithm to obtain real galaxy systems in deep spectroscopic/photometric redshift surveys.
To achieve this goal, we have constructed a synthetic galaxy catalogue using one of the largest
simulated galaxy samples available at the present, the semi-analytical galaxies built by \cite{guo11} 
on top of the Millennium Simulation I. 
We note that adopting any specific semi-analytical model could introduce a dependence of the 
results on the particular set of parameters and physical processes that were used in the model 
construction. Nevertheless, analysing the differences caused by using different semi-analytical 
models is beyond the scope of this work

To build a light-cone mock catalogue we use the information available at different 
evolutionary stages in order to reproduce temporal galaxy evolution. 
We have applied several recipes into the mock catalogue construction procedure to avoid 
different problems that arise from the construction technique,
like missing/duplicate galaxies and repetition of structures in the survey. 
The mock catalogue is tailored to the future J-PAS apparent magnitude limit and photometric band.
The technique to compute photometric redshifts for each mock galaxy is also the same that 
will be applied to that future photometric all-sky survey. 
The resulting light-cone mock catalogue comprises roughly 800,000 galaxies down to an 
observer frame apparent magnitude of $23$ in the SDSS $i$-band, 
with a median redshift of $0.72$ and a maximum of $1.5$ within a solid angle of 17.6 $deg^2$. 

Firstly, we sought the proper linking lengths to apply in a FoF algorithm in order to identify 
galaxy groups in a deep spectroscopic redshift survey. We analysed completeness and purity of the sample 
on the basis of a comparison member-to-member between the identified groups and a reference sample. 
The analyses of completeness and purity of the resulting sample revealed that the best identification 
is obtained when the algorithm takes into account the
variation of the galaxy luminosity function with the redshift, as well as a linear redshift dependence 
of the radial fiducial velocity in the line of sight direction. 
The best choice of the linking lengths are those that lead to a compromise between the completeness and 
the purity of the resulting sample. 
In the best scenario, we are able to identify a galaxy group sample in the spectroscopic catalogue 
that contains more than $40\%$ of highly pure groups (completely pure or with a few interlopers), 
at the same time that we are able to recover $70\%$ of highly complete groups 
(completely recovered or with only a few missing galaxies). The percentage of groups that contain at 
least part of a true group is $57\%$ (in other words, $43\%$ of the groups are completely false 
identifications), while $88\%$ of the true groups are recovered in the identification process either 
in one or several groups (only $12\%$ of the true groups are completely lost)

Secondly, using the procedure developed by \cite{liu08}, we have adapted the FoF algorithm in the 
line of sight direction into a probabilistic algorithm (pFoF) to work with photometric redshifts 
as distance estimators. Our analyses were performed in order to determine which is the proper 
probability distribution function that best describes the data and that leads to the most reliable 
group identification.On the other hand, we determine the best probability threshold that produces 
the most complete and pure sample of groups.
By comparing the spectroscopic and photometric information of the mock galaxies, we observe that
a Lorentzian probability distribution function performs better than a Gaussian function
to quantify the discrepancies between the photometric and spectroscopic redshifts. However, 
after using both distribution functions in the identification procedure for different probability
thresholds, we observe that the percentages of completely false and missing groups show
little differences as a function of the adopted distribution function. 
Adopting a compromise between the completeness and purity of the resulting sample,
we have determined that the best identification is obtained when using
a probability threshold of $70\%$ of the maximum value. The resulting sample includes 
less than $40\%$ of false identifications while it is able to recover around $60\%$
of the true groups.

We have also observed that, regardless of whether the redshifts are spectroscopic or photometric, 
the group samples are strongly improved (in terms of purity) when using only groups 
with more than 10 galaxy members. 

This work may be used to predict the number of groups that the algorithm described in this paper might 
find when applied to the future J-PAS survey. Taking into account the survey geometry, we expect
to obtain a sample of  $\sim 6,000,000$ groups with low membership ($4\leq N < 10$) and
$\sim 1,000,000$ groups with high membership ($N\geq 10$) when applying the PFOF algorithm with a 
Lorentzian probability function, an overdensity contrast that resembles the used for DM halos, 
 and a probability threshold of $70\%$, out to $z=1.2$. 
On the other hand, if we adopt a higher contour overdensity contrast assuming that galaxies are 
more concentrated than dark matter, we would obtain a galaxy group sample for the future J-PAS survey
of $\sim 4,000,000$ low membership groups and $\sim 650,000$ groups with high membership 
when applying the PFOF algorithm with a Lorentzian probability function and a probability 
threshold of $60\%$, out to $z=1.2$ (see appendix~\ref{apendB} for details).

However, it is worth noticing that the choice of the probability threshold should be 
done according to the final purpose of the group sample: if the obtained groups will be 
used as proxies for other group searching algorithms, then one may choose a low probability threshold 
which would imply a sample with a high completeness level (low purity); if the groups will be 
used for performing analyses of group properties, then it is better to choose a high probability 
threshold which would imply a high purity level (low completeness).

Finally, given that our criteria to define purity and completeness of groups are
very detailed and restrictive, we were able to assess the different types of groups that contribute 
to the resulting identified sample. Using more relaxed criteria to define pure and complete groups,  
as well as different reference samples, could lead to higher percentages compared to those 
found in this work.   

During the latest stages of this work, we have become aware of the existence of a recently submitted 
work by \cite{jian13}. In that work, the authors have performed a similar analysis as the one 
presented here, i.e., using the \cite{liu08} adaptation of the FoF algorithm to identify galaxy groups but 
in the Pan-STARRS1 Medium Deep Surveys. Even though both works pursuit similar objectives 
about assessing the reliability of galaxy group identification in photometric redshift surveys, the 
approaches adopted in both works are quite different. For instance, the semi-analytical galaxies, 
the procedures for determining the proper linking length parameters, 
the reference samples, the criteria to compute purity and completeness of identified 
groups as well as the way to compute the photometric redshifts in 
mock catalogues are some of the points where the two works clearly differ. 
Although it is difficult to perform a fair comparison among the two works, we note 
that our values of purity and completeness are overall consistent with those obtained by 
\citeauthor{jian13}.

\begin{acknowledgements}
We thank Manuel Merch\'an, Mario Sgr\'o and Ra\'ul Angulo for useful discussions and suggestions.
The Millennium Simulation databases used in this paper and the web application providing online 
access to them were constructed as part of the activities of the German Astrophysical 
Virtual Observatory (GAVO).
We thank Qi Guo for allowing public access for the outputs of her very impressive
semi-analytical model of galaxy formation.
This work has been partially supported by Consejo Nacional de Investigaciones Cient\'\i ficas y 
T\'ecnicas de la Rep\'ublica Argentina (CONICET, PIP2011/2013 11220100100336), Secretar\'\i a de 
Ciencia y Tecnolog\'\i a de la Universidad de C\'ordoba (SeCyT) and Funda\c c\~ao de Amparo \`a 
Pesquisa do Estado do S\~ao Paulo (FAPESP), through grants 2011/50471-4 and 2011/50002-4. CMdO 
acknowleges support of FAPESP (grant \#2006/56213-9) and Conselho Nacional de Pesquisas (CNPq). 
AZ and EDG wish to thank the IAG staff for the hospitality during the extended visit, when part 
of this work was done.
\end{acknowledgements}

\bibliography{refs}

\appendix

\section{Testing the mock catalogue: non-interpolated galaxy positions and velocities}
\label{apendA}
We performed an additional test using a different galaxy lightcone mock catalogue 
constructed using the original galaxy positions and peculiar velocities 
obtained from each simulation snapshot 
(hereafter, Non-Interpolated Positions and Velocities, NIPV).

The new mock catalogue comprises $6,756,931$ galaxies with absolute magnitudes brighter than
$-16.4$ up to $z=1.5$, i.e, $0.01\%$ more galaxies than in the 
interpolated positions and velocities (hereafter, IPV) mock catalogue. 

Following the procedure described in Sect.~\ref{reference-samples},
we identified a new \emph{reference sample} for the NIPV mock catalogue. 
A comparison between the resulting group sample for the NIPV
mock catalogue and the original IPV mock catalogue is shown in Table~\ref{a1}.

From the Table, it can be seen that the new reference group sample is 
only $2.5\%$ larger than the IPV group sample, and comprises $3.7\%$ more galaxies. 
To investigate intrinsic differences 
among the groups of both reference samples we performed a comparison member by member.
If we use the IPV group sample as reference, our comparison shows that the $95\%$ of the 
NIPV groups are directly correlated with the IPV group sample, while only $5\%$ of NIPV 
groups are intrinsically different. On the other hand, 
using the NIPV group sample as reference,
$98\%$ of the IPV groups are directly correlated with the NIPV groups sample, while only
$2\%$ of IPV groups are missing in the NIPV group sample. Therefore, from this two-way
comparison, we conclude that both reference samples show a high level of 
statistical agreement.  

\begin{table}[ht]
\begin{center}
\tabcolsep 5pt
\caption{Reference groups samples identified in the IPV and NIPV mock catalogues 
\label{a1}}
\begin{tabular}{cccc}
\hline
Mock        & Total number & \multicolumn{2}{c}{Total number of groups}  \\ 
Catalogue   & of galaxies  & $4\le N < 10$ & $N\ge 10$\\
\hline
IPV  & 1,825,303 & 159,258 & 41,774 \\
NIPV & 1,893,860 & 162,763 & 43,409 \\
\hline
\end{tabular}
\end{center}
\end{table}

Nevertheless, small differences in the positions/velocities of galaxies in both
 mock catalogues could still have an impact 
on the resulting computations of purity and completeness of different group identifications 
carried out in this work. Hence, we have performed a second test in order to quantify
the impact of using a NIPV mock catalogue in the results obtained in our work.
On the NIPV mock catalogue, we performed the same procedure described in 
Sect.~\ref{zdistortiden}.
First, we use the NIPV group reference sample and compute the maximum (and second maximum) 
velocity difference of the members in the line-of-sight to their closest neighbours. 
As expected from the very good statistical agreement among the reference samples, the
values previously obtained in Sect.~\ref{zdistortiden} are also the best values 
for the NIPV group sample, i.e, $V_0=130(1+z) \ {\rm km/s}$ and $V_0=70(1+z) \ {\rm km/s}$.
Second, we reproduced the test previously performed on the volume limited 
IPV mock catalogue to analyse the effect of distortions in redshift space,
by performing an identification of groups in redshift space on the volume limited 
NIPV mock catalogue using four different linking length parameters in the 
line-of-sight direction: $130$, $130(1+z)$, $70$ and $70(1+z)$. 
The percentages of purity and completeness of groups split into six categories 
obtained for the NIPV groups samples are shown in Table~\ref{a2}. 
For a direct comparison, we also included the previous findings associated with 
the IPV samples. 
 
\begin{table}
\begin{center}
\tabcolsep 5pt
\caption{Total percentages of purity and completeness of groups identified in  
the IPV/NIPV mock catalogues \label{a2}}
\begin{tabular}{ccccc}
\hline
Class      & \multicolumn{4}{c}{Redshift Space - $V_0$} \\ 
\hline
   & $130$ &  $130(1+z)$ & $70$  & $70(1+z)$ \\
\hline
P1 & 42/42 & 35/35 & 49/47 & 42/41 \\
P2 & 21/21 & 21/21 & 20/20 & 21/21 \\   
P3 &  6/7  &  5/5  &  8/9  &  6/6  \\
P4 & 12/13 & 11/12 & 11/12 & 12/13 \\
P5 &  2/1  &  1/1  &  1/2  &  1/1  \\
P6 & 17/16 & 27/26 & 11/10 & 18/18 \\
\hline
C1 & 14/12 & 48/44 &  3/3  & 17/15 \\
C2 & 16/14 & 25/26 &  5/4  & 18/16 \\   
C3 &  6/6  &  8/9  &  2/1  &  7/7  \\
C4 & 21/22 &  8/9  & 19/18 & 21/22 \\
C5 &  6/7  &  1/1  &  7/7  &  5/6  \\
C6 & 37/39 & 10/11 & 64/67 & 32/34 \\
\hline
\end{tabular}
\end{center}
\end{table}

From the comparison with the values obtained for the IPV group samples, it is quite clear
that identifying groups on a NIPV mock catalogue does not introduce statistical 
significant differences in the corresponding percentages of purity and completeness 
of groups. Therefore, we conclude that the adopted IPV mock catalogue used throughout 
our work does not introduce any particular bias in our results.

\section{Groups identified with higher contour overdensity contrast} 
\label{apendB}
Properties of groups of galaxies depend sensitively on the algorithm 
for group selection. In the past, groups of galaxies have been identified 
in observational catalogues with FoF linking lengths corresponding to different 
contour overdensity contrasts: 20 \citep{geller83}, 
80 \citep{ramella89,merchan02,merchan05}, 200 \citep{zandivarez11}
or 365 \citep{berlind06}. 
According to \cite{knebe13}, it must be stressed that there is no right or wrong way; 
users of halo finder catalogues just need to be aware that several alternative 
definitions exist and which one of these has been used, especially when 
computing masses and other group properties. 

In this section we apply a different contour overdensity contrast to identify groups 
in the mock galaxy catalogues. Some authors argued that since galaxies are 
more concentrated than dark matter, a higher contour overdensity contrast should be used 
\citep{eke04,berlind06}. 
Therefore, using these works we modify the empirical contour overdensity contrast of 
\cite{courtin11}, shown in Eq.~\ref{rho}, by lowering the original linking length parameter $b_{0}$ 
from 0.2 to 0.14.
Note that at redshift z=0, this formula leads to $ \delta \rho/\rho=433$ 
compared to  $\delta \rho/\rho=148 $ that has been used in the main 
body of this work. As stated previously in Sect~\ref{reference-samples}, 
it should be reminded that the redshift dependence only introduce
a variation of the linking length parameter of $\sim8\%$ in the whole redshift range.

With the aim of analysing the effect of a different overdensity in the performance of 
the group finder, we repeated all the stages of this work for this new identification. 
The new \emph{reference sample} identified in real space comprises $164,580$ groups 
with more than 4 members. This sample has $18\%$ less groups than the sample identified 
with a lower contour overdensity contrast. 

\begin{table}
\begin{center}
\tabcolsep 5pt
\caption{Total percentages of purity and completeness of groups (with $z= 0 -1.2$) identified 
in different mock galaxy samples \label{percentage2}}
\begin{tabular}{cccccccc}
\hline
{\small Class} & \multicolumn{2}{c}{\small Flux limited} &  \multicolumn{4}{c}{\small Redshift - $V_0$} & {\small Sp-mock} \\ 
      & {\tiny LF-f} & {\tiny LF-v}  &  {\tiny $130$} &  {\tiny $130(1+z)$} & {\tiny $70$}  & {\tiny $70(1+z)$} & {\tiny $130(1+z)$}\\
\hline
P1   & 40 & 52 & 45 & 38 & 52 & 44 & 23 \\
P2   & 23 & 22 & 22 & 23 & 21 & 22 & 20 \\   
P3   & \ 1 & \ 1 & \ 5 & \ 4 & \ 6 & \ 5 & \ 2 \\
P4   & \ 7 & \ 5 & 11 & 11 & 10 & 11 & 13 \\
P5   & \ 0 & \ 0 & \ 1 & \ 1 & \ 1 & \ 1 & \ 1 \\
P6   & 29 & 20   & 16 & 23 & 10 & 17 & 41 \\
\hline
C1   & 89 &  82 & 12 & 42 & \ 3 & 14 & 38 \\
C2   & \ 7 & 11 & 13 & 25 & \ 4 & 15 & 26 \\
C3   & \ 2 & \ 2 & \ 4 & \ 7 & \ 1 & \ 4 & \ 6 \\
C4   & \ 1 & \ 2 & 21 & 11 & 16 & 21 & 12 \\
C5   & \ 0 & \ 0 & \ 5 & \ 1 & \ 5 & \ 5 & \ 2 \\
C6   & \ 1 & \ 3 & 45 & 14 & 71 & 41 & 16 \\
\hline
\end{tabular}
\end{center}
\end{table}

\begin{table}
\begin{center}
\tabcolsep 5pt
\caption{Total percentages of purity and completeness of groups identified in a mock galaxy catalogue with realistic photometric redshifts \label{total_PC_photoz2}}
\begin{tabular}{ccccccccc}
\hline
Class & & \multicolumn{7}{c}{BPZ - Lorentzian - $P_{th}$ }  \\ 
      & & 30 & 60  &  70 &  80 & 90 & 95 & 99\\
\hline
P1   & & \ 5 & \ 9& 11 & 13 & 16 & 20 & 34  \\
P2   & & 11 & 16 & 19 & 21 & 24 & 27 & 26 \\
P3   & & \ 2 & \ 4 & \ 4 & \ 5 & \ 7 & \ 8 & \ 9 \\
P4   & & 25 & 26 & 27 & 26 & 26 & 24 & 18 \\
P5   & & \ 4 & \ 4 &  \ 3 & \ 3 & \ 3 & \ 3 &  \ 2 \\
P6   & & 53 & 41   & 36 & 32 & 24 & 18 & 11 \\
\hline
C1   & & 11 &  \ 5 & \ 3 & \ 2 & \ 1 & \ 0 & \ 0 \\
C2   & & 25 & 14 & 10 & \ 6 & \ 3 & \ 1 & \ 0 \\
C3   & & 10 & \ 3 & \ 3 & \ 2 & \ 1 & \ 0 & \ 0 \\
C4   & & 25 & 29 & 28 & 25 & 19 & 13 & \ 3 \\
C5   & & \ 4 & \ 5 & \ 5 & \ 5 & \ 5 & \ 4 & \ 1\\
C6   & & 25 & 44 & 51 & 60 & 71 & 82 & 96  \\
\hline
\end{tabular}
\end{center}
\end{table}

\begin{figure}
\begin{center}
\includegraphics[width=85mm]{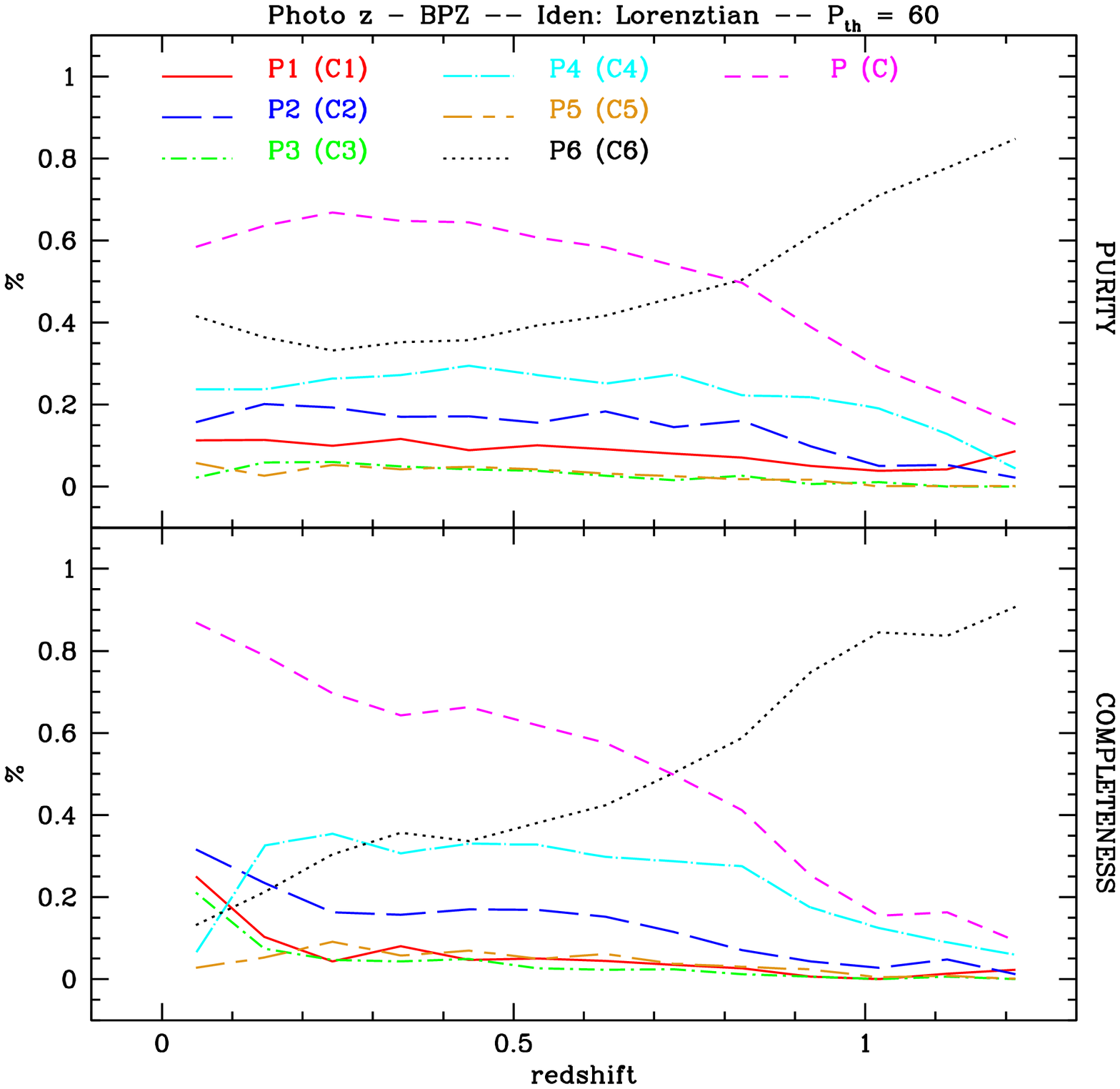}
\caption{\emph{Top panel}: Percentage of photometric groups split
into the six categories of purity as a function of redshift. 
\emph{Bottom panel}: Percentage of \emph{restricted-reference} split into the 
six categories of completeness as a function of redshifts
}
\label{pctotz2}
\end{center}
\end{figure}
\begin{figure}
\begin{center}
\includegraphics[width=85mm]{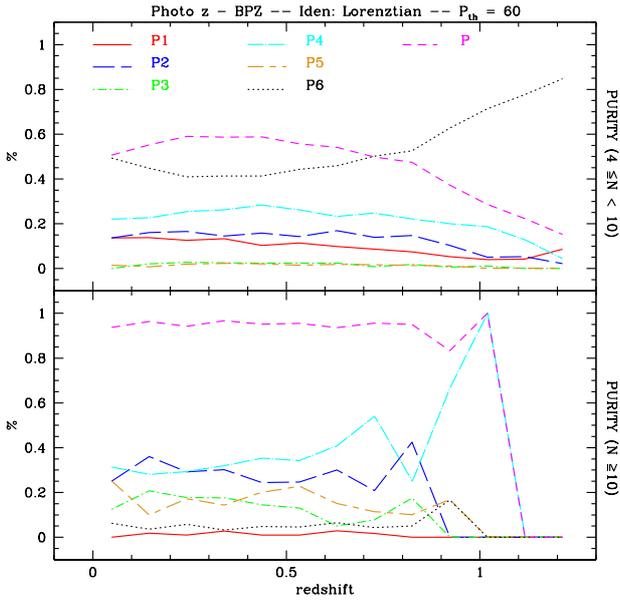}
\caption{Purity of groups identified in a realistic photometric mock galaxy sample.
\emph{Top panel}: Percentage of low membership photometric groups split
into the six categories of purity as a function of redshift. 
\emph{Bottom panel}: Percentage of high membership photometric groups split
into the six categories of purity as a function of redshift. 
}
\label{pctotz_num2}
\end{center}
\end{figure}

We performed the tests of the FoF algorithm against redshift space distortions 
and flux limit. The appropriate linking length in the line of sight direction 
was determined in the same way as we did before: by measuring the maximum separation 
in the distorted radial direction to the closest neighbour (and second closest neighbour). 
We found no differences in the result. In Table~\ref{percentage2}, 
we quote the percentages of groups split into the different categories 
of purity and completeness. By comparing with Table~\ref{percentage}, 
it can be seen that there is no change in the behaviour 
of the group finder. The samples obtained with a higher contour overdensity 
contrast exhibit the same purity and completeness as the sample obtained 
with a lower overdensity contrast. 
The final spectroscopic sample of galaxy groups comprises $16,336$ groups with more 
than 4 members, i.e., it has $30\%$ less groups than the sample obtained with the 
higher overdensity.

We also tested the algorithm in photometric redshift space and determined the 
best probability threshold in terms of purity and completeness.
Table~\ref{total_PC_photoz2} shows 
the percentage of groups into the different categories of purity and completeness 
for different probability thresholds. It can be seen that 
the best compromise between purity and completeness is reached when adopting a 
probability threshold lower than in the identification with lower overdensity contrast. 
In this case, the best choice in terms of purity and completeness is $P_{th}=60\%$ which 
produces a sample of $10,740$ groups with more than 4 members, 
which has $31\%$ less groups than the best sample identified with the higher 
overdensity contrast and $P_{th}=70\%$. In this sample, the percentage of false groups 
is $41\%$ while the percentage of missing groups is $44\%$.

In Fig.~\ref{pctotz2} we show the variation of the six categories of purity and 
completeness for the sample identified with $P_{th}=60\%$ as a function of redshift. 
This figure can be directly compared to Fig.~\ref{pctotz} to observe that changing the contour 
overdensity contrast does not introduce major differences in the purity/completeness of 
the resulting sample provided the probability threshold is also changed.

Finally, in Fig.~\ref{pctotz_num2} we also show the behaviour of the purity of the sample when 
groups, identified with $P_{th}=60\%$, are split into low (9,323 groups) and high (1417 groups) 
membership. It can be seen that the low membership sample 
introduces the highest percentage of false identifications (P6), and therefore we recommend 
to avoid using those groups when performing statistical analyses of the properties of these groups.
This result is very similar to the one previously shown in Fig.~\ref{pctotz_num} using a lower
overdensity contrast in the identification process.   

\end{document}